\documentclass[aps,prc,twocolumn,superscriptaddress,nofootinbib,tightenlines,
preprintnumbers,showkeys,floatfix]{revtex4-2}

\usepackage[utf8]{inputenc}
\usepackage{CJK}
\usepackage[english]{babel}
\usepackage{amssymb,amsthm,amsmath,amstext,amsbsy,amsopn,mathrsfs}
\usepackage{bbm}
\usepackage{nicefrac}
\usepackage{suffix}
\usepackage{array}
\usepackage{slashed}
\usepackage{leftidx}
\usepackage{graphicx}
\usepackage{hyperref}
\usepackage{environ}
\usepackage{mathtools}
\usepackage{xspace}
\usepackage{xcolor}
\usepackage{pgfplotstable}
\usepackage{scalerel,stackengine}
\usepackage{booktabs,colortbl}
\usepackage{physics}
\usepackage{braket}
\usepackage{isotope}
\usepackage{MnSymbol}

\newcommand{\ie}{\textit{i.e.}\xspace}
\newcommand{\eg}{\textit{e.g.}\xspace}
\newcommand{\cf}{\textit{cf.}\xspace}
\newcommand{\apriori}{\textit{a priori}\xspace}
\WithSuffix\newcommand\apriori*{\textit{a-priori}\xspace}

\newcommand{\abinitio}{\textit{ab initio}\xspace}

\newcommand{\fm}{\ensuremath{\text{fm}}}

\renewcommand{\vec}[1]{\mathbf{#1}}

\newcommand{\vecp}{\mathbf{p}}

\newcommand{\vecq}{\mathbf{q}}

\newcommand{\vecK}{\mathbf{K}}

\newcommand{\vsigma}{\boldsymbol{\sigma}}

\newcommand{\vdelta}{\delta^{(3)}}

\newcommand{\ii}{\mathrm{i}}
\newcommand{\ee}{\mathrm{e}}

\newcommand{\OO}{\mathcal{O}}

\newcommand{\YY}{Y}
\newcommand{\YYY}{\scalebox{1.33}{\ensuremath{\mathcal{Y}}}}

\newcommand{\CG}[6]{\ensuremath{C^{#5#6}_{#1#2,#3#4}}}
\WithSuffix\newcommand\CG*[3]{\ensuremath{C^{#3}_{#1,#2}}}

\newcommand{\SixJ}[6]{\left\{\!\!%
\begin{array}{ccc}%
{#1} & {#2} & {#3}\\[0.2em]%
{#4} & {#5} & {#6}%
\end{array}%
\!\!\right\}%
}

\newcommand{\NineJ}[9]{\left\{\!\!%
\begin{array}{ccc}%
{#1} & {#2} & {#3}\\[0.2em]%
{#4} & {#5} & {#6}\\[0.2em]%
{#7} & {#8} & {#9}%
\end{array}%
\!\!\right\}%
}

\newcommand{\couple}[3]{\left({#1}{#2}\right)\!{#3}}

\newcommand{\nnch}[3]{\ensuremath{{}^{#1}\mkern-3mu#2_{#3}}}

\stackMath
\newcommand\reallywidehat[1]{%
\savestack{\tmpbox}{\stretchto{%
  \scaleto{%
    \scalerel*[\widthof{\ensuremath{#1}}]{\kern-.6pt\bigwedge\kern-.6pt}%
    {\rule[-\textheight/2]{1ex}{\textheight}}
  }{\textheight}%
}{0.5ex}}%
\stackon[1pt]{#1}{\tmpbox}%
}

\newcommand{\MN}{M_N}
\newcommand{\Md}{M_d}

\newcommand{\ec}{e}
\newcommand{\Bd}{B_d}

\newcommand{\Mhi}{M_{\text{hi}}}

\newcommand*\rvec[1]%
{\ensuremath{\overset{\smash{\raisebox{-1.5pt}{\tiny$\rightarrow$}}}{#1}}}
\newcommand*\lvec[1]%
{\ensuremath{\overset{\smash{\raisebox{-1.5pt}{\tiny$\leftarrow$}}}{#1}}}

\newcommand{\MeV}{\ensuremath{\text{MeV}}}

\NewEnviron{subalign}[1][]{%
\begin{subequations}\begin{align}
  \BODY
\end{align}\label{#1}\end{subequations}
}

\NewEnviron{spliteq}{%
\begin{equation}\begin{split}
  \BODY
\end{split}\end{equation}
}

\NewEnviron{mleq}{%
\begin{equation}
  \BODY
\end{equation}
}

\newcommand{\ALIT}{\mathcal{A}}
\newcommand{\BLIT}{\mathcal{B}}

\definecolor{forest}{RGB}{34, 139, 34}

\newcommand{\mhi}{\Mhi}

\begin{document}

\title{Perturbative EFT calculation of the deuteron longitudinal response
function}

\begin{CJK*}{UTF8}{}
\CJKfamily{gbsn}

\author{Andrew J.~Andis}
\email{ajandis@ncsu.edu}
\affiliation{Department of Physics and Astronomy,
North Carolina State University,
Raleigh, North Carolina 27695, USA}

\author{Songlin Lyu (吕松林)}
\affiliation{Dipartimento di Matematica e Fisica,
Universit\`a degli Studi della Campania ``Luigi Vanvitelli'',
viale Abramo Lincoln 5 - I-81100 Caserta, Italy}
\affiliation{Istituto Nazionale di Fisica Nucleare, \\
Complesso Universitario di Monte  S. Angelo, Via Cintia - I-80126 Napoli, Italy}

\author{Bingwei Long (龙炳蔚)}
\affiliation{College of Physics, Sichuan University,
Chengdu, Sichuan 610065, China}
\affiliation{Southern Center for Nuclear-Science Theory (SCNT),
Institute of Modern Physics, Chinese Academy of Sciences,
Huizhou 516000, Guangdong, China}

\author{Sebastian K\"onig}
\email{skoenig@ncsu.edu}
\affiliation{Department of Physics, North Carolina State University,
Raleigh, North Carolina 27695, USA}

\begin{abstract}
 In this work, we study the longitudinal response function of the deuteron
 up to next-to-next-to-leading order in chiral effective field theory (Chiral
 EFT).
 We use an approach that maintains exact renormalization group (RG) invariance
 at each order of the EFT expansion by treating all subleading corrections in
 perturbation theory.
 To that end, we extent the Lorentz Integral Transform (LIT) method to allow for
 such a perturbative treatment.
 In doing so, we further develop the existing work on strictly RG invariant
 Chiral EFT, which has so far focused primarily on binding energies and static
 properties, to inelastic processes.
 We carefully analyze the convergence properties of the theory and find good
 agreement with available experimental data.
 Our findings provide the foundation for similar studies of inelastic processes
 in a range of nuclei, based on perturbatively renormalized EFT schemes.
\end{abstract}

\maketitle

\end{CJK*}

\section{Introduction}
\label{sec:Introduction}

Nuclear effective field theories (EFTs) are a cornerstone of modern low-energy
nuclear theory because they enable calculations of atomic nuclei to be
described as fully dynamical few- and many-body systems of protons and neutrons
with interactions between the nucleons derived as a systematic approximation of
Quantum Chromodynamics (QCD).
The approximation takes the form of an expansion in powers of $Q/\mhi$, where
$Q$ is the typical size of momenta of the particles involved in the processes of
interest and $\mhi$ is the breakdown scale of the EFT that is associated with,
\eg, the mass of the heavy particles or the energies of some highly excited
states that are ``integrated out.''
Although the physics associated with these scales does not enter
explicitly, their effects on low-energy observables are encoded in the
coupling strengths of various contact (zero-range) interactions within the EFT.

Of the various EFTs relevant for low-energy nuclear structure and reactions,
most recently reviewed in Ref.~\cite{Hammer:2019poc}, the so-called Chiral EFT
(ChEFT) is generally assumed to be applicable for describing a significant
fraction of the chart of nuclides.
Based on the assumption that the (approximate) chiral symmetry of QCD, $SU(2)_L
\times SU(2)_R$, is spontaneously broken, one can write down a Chiral
Lagrangian that manifests this symmetry in terms of the pionic and nucleonic
degrees of freedom~\cite{Weinberg:1990rz, Weinberg:1991um, Ordonez:1992xp,
Weinberg:1992yk, vanKolck:1993}.
This approach then describes the strong interactions between nucleons, and also
electroweak processes~\cite{Rho:1990cf, Park:1993jf, Park:1995pn,
Kolling:2009iq, Kolling:2011mt, Pastore:2011ip, Baroni:2015uza, Krebs:2020pii}.
Even extensions beyond the Standard Model can be included in the
Chiral Lagrangian; see, for example, Refs.~\cite{Hoferichter:2015ipa,
Cirigliano:2018yza, Oosterhof:2019dlo, Liu:2022cfd}.
However, the EFT Lagrangian has in principle an infinite number of operators,
and therefore one needs an organization scheme, the so-called power counting,
to decide which operators in the Lagrangian are needed for a
desired target precision of observables calculated from the EFT.

In the most simple scheme, referred to as naive dimensional
analysis (NDA), one counts the dimension of operators in the Lagrangian, built
from a combination of fields, and derivatives acting on them (or momenta
multiplying them, if the theory is formulated in momentum space), and then one
places powers of $\mhi$ in the denominators to fix the overall dimension of the
term with a dimensionless coupling strength.
Although popular in applications of ChEFT to nuclear physics, NDA can be
unreliable because other soft mass scales, such as the pion decay constant
$f_\pi$, are present in the theory and therefore an operator may be related
to $Q/f_\pi$ rather than the suppressed ratio $Q/\mhi$.
It is possible to use Renormalization Group (RG) invariance to identify
enhancement of interactions terms relative to NDA \emph{before} one even
confronts the EFT with experimental data.
For example, the RG constraint---\ie, that observables are insensitive to an
arbitrarily chosen ultra-violet momentum cutoff $\Lambda$ used to regularize
the otherwise ill-defined contact interactions in the Lagrangian---results in
some of the nucleon-nucleon ($NN$) contact interactions being promoted to
lower orders than they would appear according to NDA~\cite{Nogga:2005hy,
PavonValderrama:2011fcz, Long:2011qx}.
Similar effects occur in the so-called Pionless EFT, where the large
nucleon-nucleon S-wave scattering lengths lead to shallow bound and virtual
$NN$ states that eventually give rise to relevant low-energy scales.
These and other effects are reviewed in detail in Ref.~\cite{Hammer:2019poc}.

Reference~\cite{Shi:2022blm} extended the exploration of ChEFT with
perturbative RG invariance to electromagnetic current operators by studying
static properties (form factors) of the deuteron with use of perturbative
RG-invariant chiral potentials developed in Refs.~\cite{Long:2011xw,
Long:2012ve, Wu:2018lai}.
An important finding of this study is that, similar to what has been found for
the chiral expansion of the nuclear interaction, the power counting for
electromagnetic current operators in ChEFT needs to be modified, partially in
line with the conclusion of a previous study based on the short-distance
behavior of two-nucleon interacting wave function derived from the one-pion
exchange (OPE) potential~\cite{PavonValderrama:2014zeq}.
In this work, we extend this line of research to deuteron breakup, using
deuteron electrodisintegration as a specific example.
This inelastic reaction depends on both the initial deuteron bound state as well
as on the continuum of neutron-proton final states, thus, by focusing on this
process, we study a strictly larger range of physics than
Ref.~\cite{Shi:2022blm}.
In particular, while the initial deuteron state is determined by the interaction
in only the \nnch3S1-\nnch3D1 coupled channels, the final-state interaction
involves a number of other partial waves.

Direct calculation of dynamic observables for deuteron breakup is a feasible
problem.
However, such calculations of nuclear breakup reactions become
increasingly difficult as the number of particles in the system increases.
For this reason, we focus in this work on the Lorentz Integral Transform (LIT)
method, first introduced in Ref.~\cite{Efros:1994iq} and reviewed in
Ref.~\cite{Efros:2007nq}.
The LIT makes it possible to perform breakup calculations with methods similar
to those for bound states (albeit at the expense of working effectively with a
complex energy), which enables \abinitio many-body studies of such reactions for
a range of atomic nuclei (see Ref.~\cite{Marino:2025auh} and earlier references
therein).
A delicate inversion of the LIT is generally required to obtain the
actual response function.
Although it is not our main focus, we will review and refine an existing
approach for inverting the LIT in order to obtain the response
function.
We also note that as an alternative to the LIT it has recently been suggested
to work instead with integrated response functions, which can similarly be
obtained with bound-state methods~\cite{Griesshammer:2024twu}.

As a central part of this work, we derive the perturbative expansion of
the LIT for a two-nucleon system, which is a crucial step towards studying
nuclear breakup reaction with perturbatively renormalized EFTs.
More precisely, in this approach, only leading order (LO) is treated
nonperturbatively, with all corrections on top of that included in
distorted-wave perturbation theory.
This approach ensures that the EFT satisfies RG invariance at each
order~\cite{Hammer:2019poc}.

This paper is structured as follows.
In Sec.~\ref{sec:Theory}, we discuss in detail the theoretical framework for
our calculation, including the LIT for the particular case of deuteron
electrodisintegration and the perturbatively expanded, momentum-space
representation.
In this section we also briefly review the chiral interaction and current
operators that we are using.
Following this, we present selected results in Sec.~\ref{sec:Results} before we
conclude with a summary and outlook in Sec.~\ref{sec:Conclusion}.
Several more technical details are discussed in the appendix.

\section{Theoretical framework}
\label{sec:Theory}

\subsection{Response functions for electrodisintegration}

We consider the process where an electron ($e$) scatters inelastically off
a deuteron ($d$), breaking it up into its neutron ($n$) and proton ($p$)
constituents.
The cross section for this $d(e,e)np$ process can be written
as~\cite{Forest:1966,Carlson:1997qn,Bacca:2014}
\begin{multline}
 \frac{d\sigma}{d\Omega d\omega} = \sigma_{\text{Mott}} \Bigg[
  \frac{Q^{4}}{\vecq^{4}}\mathcal{R}_{\mathrm{L}}(\omega,\vecq) \\
  \null + \Bigg(\frac{Q^{2}}{2\vecq^{2}} + \tan\frac{\theta_e}{2}\Bigg)
  \mathcal{R}_{\mathrm{T}}(\omega,\vecq)
 \Bigg] \,,
 \label{eq:cross}
\end{multline}
where $\theta_e$ is the angle of the scattered electron and $\omega$, $\vecq$
are, respectively, the energy and three-momentum of a virtual space-like
photon.
Moreover, $Q^{2} = {-}q_{\mu}^{2} = \omega^{2} - \vecq^{2}$ is the (positive)
four-momentum transfer.
Unless otherwise noted, we consider the energy and momentum components defined
in the the c.m.\ frame of the outgoing nucleons.
The prefactor $\sigma_{\text{Mott}}$ in Eq.~\eqref{eq:cross} is the Mott cross
section and is given by
\begin{equation}
 \sigma_{\text{Mott}} =
 \Bigg( \frac{\alpha \cos{\theta_e/2}}{2E \sin^2\theta_e/2} \Bigg)^2 \,,
\end{equation}
with $\alpha \approx 1/137$ the electromagnetic fine-structure constant and
$E$ the initial electron energy ($\omega = E - E'$ with $E'$ the energy of
the scattered electron).

The objects $\mathcal{R}_{\mathrm{L}}(\omega, \vecq)$ and
$\mathcal{R}_{\mathrm{T}}(\omega, \vecq)$ are the so-called longitudinal and
transverse response functions, respectively, and they can be expressed in terms
of nuclear matrix elements as follows:
\begin{subalign}[eqs:response]
\label{eq:response-L}
 \mathcal{R}_{\mathrm{L}}(\omega, \vecq)
 &= \frac13\sum_{f \neq i}\sum_{m_i, m_f}
 \abs{\braket{\Psi_{f}|J_0|\Psi_{i}}}^{2} \delta(\omega + E_{i} - E_{f})
 \,, \\
 \label{eq:response-T}
 \mathcal{R}_{\mathrm{T}}(\omega, \vecq)
 &= \frac13\sum_{f \neq i}\sum_{m_i, m_f} \sum_{\lambda=\pm 1}
 \abs{\braket{\Psi_{f}|J_{\lambda}|\Psi_{i}}}^{2}
 \delta(\omega + E_{i} - E_{f}) \,.
\end{subalign}
In these expressions, $\ket{\Psi_{i}}$ represents the initial deuteron bound
state and the $\ket{\Psi_{f}}$ are $np$ final states with energy $E_f$, with
total spin ($j_f=j_i=1$) projections $m_f$ (summed over) and $m_i$ (averaged
over), respectively.\footnote{Following the literature on the subject, we have
expressed $\mathcal{R}_{\mathrm{L}}$ and $\mathcal{R}_{\mathrm{T}}$ in terms of
a \emph{sum} over final states $f$, but it is understood that for transitions
to continuum states (which for the deuteron without bound excited states is in
fact the only possibility) this sum should strictly be an integral.}
$J_0$ and $J_{\lambda}$ are the charge-density and current operators in the
one-photon-exchange approximation,
respectively~\cite{Forest:1966,Carlson:1997qn,Bacca:2014}, which induce the
transitions from initial to final states, and the delta functions in
Eqs.~\eqref{eqs:response} ensure conservation of energy.
Elastic transitions ($f=i$) are explicitly excluded in the definitions of the
response functions.
For the remainder of this work, we will only consider the longitudinal response
function because this already involves a significant fraction of the physics we
are interested in, while avoiding the technical complexity of dealing with
transverse current operators.
We will also use the simpler notation $J_0 = \rho$ for the charge-density
operator.

\subsection{The Lorentz Integral Transform}
\label{sec:TheLIT}

The LIT of the longitudinal response function can be introduced as a convolution
with a Lorentzian kernel~\cite{Efros:1994iq,Efros:2007nq}:
\begin{equation}
\label{eq:LIT}
 \Phi(\sigma)
 = \int_{\omega_{\text{th}}}^{\infty} \dd\omega \,
 \tilde{\mathcal{R}}_{\mathrm{L}}(\omega)L(\omega,\sigma) \,.
\end{equation}
Here $\sigma$ is a complex parameter that in the following we mostly write out
explicitly as $\sigma = \sigma_R + \ii\sigma_I$.
The integral in Eq.~\eqref{eq:LIT} starts at the threshold energy
$\omega_{\text{th}}$, which is the transferred energy at which the breakup
becomes possible, and extends up to infinity.
The Lorentzian kernel is given by
\begin{equation}
 L(\omega,\sigma) \equiv L(\omega, \sigma_R; \sigma_I) \,
 = \frac{1}{(\omega - \sigma_R)^{2} + \sigma_I^{2}} \,,
\end{equation}
and from this form it becomes obvious that Eq.~\eqref{eq:LIT} is indeed a
convolution.
The LIT $\Phi(\sigma) \equiv \Phi(\sigma_R;\sigma_I)$ is generally considered as
a function of $\sigma_R$ with a parametric dependence on $\sigma_I$ that
characterizes the ``width,'' \ie, the resolution, of the transform.

Using this kernel, we now follow \textcite{Martinelli:1995vn} (who performed an
analogous derivation for three nucleons using the Faddeev formalism) and derive
an abstract operator equation for the LIT.
Inserting the definition of the longitudinal response function,
Eq.~\eqref{eq:response-L}, into Eq.~\eqref{eq:LIT}, averaging over the
deuteron spin projection (as discussed above), and summing over final states
(including their spin projections $m_f$), gives
\begin{equation}
\label{eq:LIT-sum-mimf}
 \Phi(\sigma_{R};\sigma_{I})
 = \frac13\sum_{f \neq i}\sum_{m_i, m_f} \,\,
  \frac{\big|\braket{\Psi_{f}|\rho|\Psi_{i}}\big|^{2}}
  {\sigma_{I}^{2} + (E_{f} - E_{i} - \sigma_{R})^2} \,.
\end{equation}
At this point, we expand the charge density operator into multipoles,
writing\footnote{%
Note that our convention for the multipoles differs from the more commonly used
form for Coulomb multipoles (discussed, for example, in Ref.~\cite{Bacca:2005}
(and \cf~also Refs.~\cite{Efros:1993xy,Efros:1994iq}) by a factor
$\sqrt{4\pi}\,\ii^L$.}
\begin{equation}
 \rho(\vecq) = \sum_{L} \sqrt{\hat{L}} \rho_{L}(q) \,,
 \label{eq:rho-L}
\end{equation}
where the integer $L \geqslant 0$ denotes the multipole and $\hat{L} = 2L+1$.
We use a coordinate system where the momentum $\vecq$ of the photon is aligned
along the $z$-axis and therefore there are no explicit spherical harmonics in
Eq.~\eqref{eq:rho-L} and the $\rho_L$ appearing on the right-hand side
are functions of $q = \abs{\vecq}$ only; see Ref.~\cite{Bacca:2005} for more
details.
In general, by the Wigner-Eckart theorem, the multipole operators $\rho_{L}(q)$
between two nucleon states can be written as
\begin{equation}
 \braket{\Psi_{f}|\rho_L|\Psi_{i}}
 = \CG{j_f}{m_f}{L}{M}{j_i}{m_i} \braket{\Psi_{f}||\rho_L||\Psi_{i}} \,,
\label{eq:Psi-fi-WE}
\end{equation}
\ie, as a product of Clebsch-Gordan coefficients and
reduced matrix elements.
$M$ here is the projection quantum number associated with $L$, and by our
choice of coordinate system only $M=0$ contributes; see Eqs.~\eqref{eq:rho-0-jm}
and~\eqref{eq:rho-2-jm} in the Appendix for the precise expressions for the
current operator that include these Clebsch-Gordan coefficients.
Finally, while in principle the sum over $L$ runs up to infinity, in practice
one finds that convergence can be reached by truncating the sum at some $L =
L_{\text{max}}$.

Inserting Eq.~\eqref{eq:Psi-fi-WE} into Eq.~\eqref{eq:LIT-sum-mimf} allows us to
explicitly evaluate the sum over spin projections of the squared matrix element:
\begin{equation}
 \sum_{m_i, m_f}
 \CG{j_f}{m_f}{L}{M}{j_i}{m_i}
 \CG{j_f}{m_f}{L'}{M'}{j_i}{m_i}
 = \frac{2j_i + 1}{2L + 1} \delta_{LL'} \delta_{MM'} \,.
\end{equation}
The numerator $2j_i + 1 = 3$ cancels against the $1/3$ from averaging, while
the denominator can be canceled against the $\sqrt{\hat{L}\hat{L'}}$ arising
form Eq.~\eqref{eq:rho-L}.
Overall, we find
\begin{equation}
 \Phi(\sigma_{R};\sigma_{I}) = \sum_L \Phi_L(\sigma_{R};\sigma_{I}) \,,
\label{eq:Phi-L}
\end{equation}
and we can further rewrite each individual $\Phi_L(\sigma_{R};\sigma_{I})$:
\begin{widetext}
\begin{spliteq}
\label{eq:LIT-sum}
 \Phi_L(\sigma_{R};\sigma_{I})
 &= \sum_{f \neq i} \,\,
  \frac{\big|\braket{\Psi_{f}||\rho_L||\Psi_{i}}\big||^{2}}
  {\sigma_{I}^{2} + (E_{f} - E_{i} - \sigma_{R})^2} \\
 &= \sum_{f \neq i} \,\,
  \frac{\braket{\Psi_{f}||\rho_L||\Psi_{i} }
  \braket{\Psi_{i}||\rho^{\dagger}_L ||\Psi_{f}}}
  {(E_{f} - E_{i} - \sigma_R + \ii\sigma_I)
  (E_{f} - E_{i} - \sigma_R - \ii\sigma_I)} \\
 &= \sum_{f \neq i}
  \braket{
   \Psi_{i}||\rho^{\dagger}_L(H-E_{i}-\sigma_R-\ii\sigma_I)^{-1} ||\Psi_{f}
  }
  \braket{
   \Psi_{f}||(H-E_{i}-\sigma_R+\ii\sigma_I)^{-1} \rho_L||\Psi_{i}
  } \,.
\end{spliteq}
\end{widetext}
In the last line we have used the fact that the final states satisfy the
Schr{\"o}dinger equation $H\ket{\Psi_f} = E_f \ket{\Psi_f}$, and we
swapped the order of the matrix elements to make the following steps more
obvious.
Finally, we note that of course there is a multipole expansion
like Eq.~\eqref{eq:Phi-L} directly for the response function, and
Eq.~\eqref{eq:LIT-sum} can be obtained from applying the LIT~\eqref{eq:LIT}
separately to each term in that expansion.

In the last line of Eq.~\eqref{eq:LIT-sum}, the sum runs over all possible
final states \emph{except} $\ket{\Psi_{i}}$, \ie, the ground-to-ground
state transition is explicitly excluded.
Further following Ref.~\cite{Martinelli:1995vn}, we can extend the sum to
include this term and subtract it at the end to compensate, noting that it can
be written in terms of the elastic form factor.
This allows us to identify the sum over all final states with the identity, by
means of completeness, and to arrive at
\begin{widetext}
\begin{equation}
\label{eq:lit_der}
 \Phi_L(\sigma_{I},\sigma_{R})
 = \braket{\Psi_{i}||\rho^{\dagger}_L (H - E_{i} - \sigma_R - \ii\sigma_I)^{-1}
  (H - E_{i} - \sigma_R + \ii\sigma_I)^{-1} \rho_L ||\Psi_{i}}
  - \frac{\abs{\braket{\Psi_{i}||\rho_L||\Psi_{i}}}^{2}}{\sigma_R^2+\sigma_I^2}
 \,,
\end{equation}
\end{widetext}
From this form one can see that the first term in
Eq.~\eqref{eq:LIT-sum} can be written as the squared norm of a state
$\ket{\Phi_{i}}$ that satisfies the LIT equation
\begin{equation}
 \label{eq:LIT-inteq}
 (H - E_{i} - \sigma_R + \ii\sigma_I) \ket{\Phi_{L}} = \rho_L \ket{\Psi_{i}} \,,
\end{equation}
where $\rho_L$ is understood to ultimately yield a reduced matrix element,
\ie, the Clebsch-Gordan coefficient from the charge density operator has already
been eliminated by summing over the total spin projections, as discussed above.
This equation is similar to the Schr\"odinger equation with an inhomogeneous
source term and a shifted, complex energy.
We can ultimately express the LIT for multipole $L$ as
\begin{equation}
\label{eq:full-transform}
 \Phi_L(\sigma_{R};\sigma_{I})
 = \braket{\Phi_L|\Phi_L}
 - \abs{\braket{\Psi_i|\Phi_L}}^{2} \,,
\end{equation}
where we have used a convenient alternative form for the second term
in Eq.~\eqref{eq:lit_der}~\cite{Martinelli:1995vn}, which we refer to as the
``elastic term.''
For $L=0$, it is proportional to the square of the deuteron form factor, and
therefore the alternative form in Eq.~\eqref{eq:full-transform}, which follows
directly from Eq.~\eqref{eq:LIT-inteq} by noting that $\ket{\Psi_{i}}$ is an
eigenstate of $H$ with energy $E_i$, can also be used to test the
LIT numerics against a direct evaluation of the form factor;
this is discussed further in Sec.~\ref{sec:FF-Sanity}.

The source term on the right-hand side of Eq.~\eqref{eq:LIT-inteq} determines
the asymptotic behavior of the solution and is the key to understanding many
features of the LIT.
In particular, together with the explicit imaginary part added to the energy,
it imposes a boundary condition that ensures that $\ket{\Phi_{m_i}}$,
when expressed in coordinate space, exhibits an exponentially decaying tail,
just like the initial bound state.
In other words, these features of the equation ensure that
${\braket{\Phi|\Phi}}$ is well defined and finite.

Moreover, we can write $H = H_0 + V$ with the free Hamiltonian $H_0$ and
rewrite Eq.~\eqref{eq:LIT-inteq} as
\begin{subalign}
 (H_{0} - E_{i} - \sigma^*)\ket{\Phi_{m_i}}
 + V\ket{\Phi} &= \rho \ket{\Psi_{m_i}} \\
 \implies
 (G_0 V - 1)\ket{\Phi_{m_i}} &= G_0 \rho \ket{\Psi_{m_i}} \,,
\label{eq:LIT-LS}
\end{subalign}
where $G_0 = G_0(E_i + \sigma^*) = (H_{0} - E_{i} - \sigma^*)^{{-}1}$
is the free Green's function evaluated at a complex energy $E_i + \sigma^*$.
The combination of operators on the left-hand side of
Eq.~\eqref{eq:LIT-inteq} is exactly what otherwise appears, for real energies, in
the Lippmann-Schwinger equation that defines the continuum scattering states.

\subsection{Partial-wave expansion}
\label{sec:PartialWaves}

In addition to using the multipole expansion for the current operator, we
also project all equations into momentum-space partial waves.
For this, we write as $\ket{p;\alpha} = \ket{p}\otimes\ket{\alpha}$ with $p$ the
$NN$ center-of-mass momentum and $\ket{\alpha} = \ket{(ls)j,t}$ collecting the
spin and isospin quantum numbers.
We will in the following often use standard spectroscopic notation to identify
particular spin/isospin channels.
We furthermore use conventions in which the completeness relation reads
\begin{equation}
 \mathbf{1} = \sum_{\alpha}\int\dd p \, p^2 \ket{p;\alpha}\bra{p;\alpha} \,.
\label{eq:PW-completeness}
\end{equation}

Equation~\eqref{eq:LIT-LS} is straightforward to project into this space, using
Eq.~\eqref{eq:PW-completeness} and the fact that $G_0$ is diagonal in both
momentum and $\alpha$.
The momentum-space partial-wave representation of the charge-density operator
multipoles, which will be introduced in more detail in Sec.~\ref{sec:Chiral}, is
discussed in Appendix~\ref{sec:CurrentPW}.
After discretizing all momenta on a quadrature mesh, Eq.~\eqref{eq:LIT-LS}
ultimately becomes an inhomogeneous linear matrix-vector equation, which is
solved with standard techniques.

In the following, we will suppress the explicit $L$ subscript for $\ket{\Phi}$,
noting that in our numerical calculation, we solve a LIT matrix-vector equation
of the form~\eqref{eq:LIT-LS} for each $L$ and ultimately sum over $L$, as
discussed in Sec.~\ref{sec:TheLIT}.
The channels $\alpha$ in which $\ket{\Phi}$ has nonzero components depend on the
multipole $L$ and on the current operator.
Different operators enter at different orders in the EFT expansion, which in
fact has to be applied to the LIT formalism overall.
This is the topic of the following subsections.

\subsection{Perturbative expansion}
\label{sec:PertLIT}

The EFT expansion gives rise to a series of potentials,
\begin{equation}
 V = V^{(0)} + V^{(1)} + V^{(2)} + \cdots \,,
\label{eq:V-series}
\end{equation}
where $V^{(0)}$ is the LO potential that is treated
nonperturbatively, and $V^{(i)}$ with $i \geqslant 1$ are perturbative
corrections.
Importantly, this expansion is to be propagated strictly to all observables
to maintain a properly renormalized theory.
For example, the energy of the initial state has an expansion of the form
\begin{equation}
 E_i = E_i^{(0)} + E_i^{(1)} + E_i^{(2)} + \cdots \,,
\label{eq:Ei-series}
\end{equation}
which follows directly from the expansion of the deuteron binding energy,
$B = B^{(0)} + B^{(1)} + B^{(2)} + \cdots$, since $E_i = {-}B$;
see Appendix~\ref{sec:DeuteronWF} for details.

We are interested in the perturbative expansion of the LIT of the form
\begin{equation}
 \Phi_L(\sigma) = \Phi_L(\sigma)^{(0)} + \Phi_L(\sigma)^{(1)}
 + \Phi_L(\sigma)^{(2)} + \cdots \,.
\end{equation}
All equations in this section are valid for a fixed multipole $L$ and it is
understand that multipole contributions are ultimately to be summed according to
Eq.~\eqref{eq:Phi-L}.
To keep the notation simple, we omit the explicit multipole subscript $L$ in the
following.

To obtain the higher-order corrections to the LIT, we expand
$\ket{\Phi} = \ket{\Phi_L}$ and $\ket{\Psi}$ in the following way
\begin{align}
 \ket{\Phi} &= \ket{\Phi^{(0)}} + \ket{\Phi^{(1)}}
 + \ket{\Phi^{(2)}} + \cdots \,, \\
 \ket{\Psi} &= \ket{\Psi^{(0)}} + \ket{\Psi^{(1)}}
 + \ket{\Psi^{(2)}} + \cdots \,,
\label{eq:Phi-Psi-exp}
\end{align}
where the latter is the standard perturbative expansion of the deuteron bound
state, reviewed briefly in Appendix~\ref{sec:DeuteronWF}.

Inserting these expansions into Eq.~\eqref{eq:full-transform} leads to two
different kinds of overlaps, which we denote as $\ALIT$ and $\BLIT$.
Up to second-order, their perturbative expansions are
\begin{subalign}
 \ALIT^{(0)} &= {\braket{\Phi^{(0)} | \Phi^{(0)}}} \,, \\
 \ALIT^{(1)} &= {\braket{\Phi^{(0)} | \Phi^{(1)}}}
  + {\braket{\Phi^{(1)} | \Phi^{(0)}}} \,, \\
 \ALIT^{(2)} &= {\braket{\Phi^{(1)} | \Phi^{(1)}}}
  + {\braket{\Phi^{(0)} | \Phi^{(2)}}}
  + {\braket{\Phi^{(2)} | \Phi^{(0)}}} \,,
\end{subalign}
and
\begin{subalign}
 \BLIT^{(0)} &= {\braket{\Psi^{(0)} | \Phi^{(0)}}} \,, \\
 \BLIT^{(1)} &= {\braket{\Psi^{(0)} | \Phi^{(1)}}}
  + {\braket{\Psi^{(1)} | \Phi^{(0)}}} \,, \\
 \BLIT^{(2)} &= {\braket{\Psi^{(1)} | \Phi^{(1)}}}
  + {\braket{\Psi^{(0)} | \Phi^{(2)}}}
  + {\braket{\Psi^{(2)} | \Phi^{(0)}}} \,,
\end{subalign}
Using these, we can now express the desired perturbative expansion of the LIT as
\begin{subalign}
 \Phi(\sigma)^{(0)}
 &= \ALIT^{(0)} - \abs*{\BLIT^{(0)}}^{2} \,, \\
 \Phi(\sigma)^{(1)} &= \ALIT^{(1)} - \BLIT^{(0)} \BLIT^{(1)*}
  - \BLIT^{(1)} \BLIT^{(0)*} \,, \\
 \Phi(\sigma)^{(2)} &= \ALIT^{(2)}
  - \abs*{\BLIT^{(1)}}^{2} - \BLIT^{(0)} \BLIT^{(2)*} \nonumber \\
  &\hspace{11em} \null  - \BLIT^{(2)} \BLIT^{(0)*} \,.
\end{subalign}

While the expansion of the deuteron state $\ket{\Psi}$ is well known, we need to
derive the equations that determine the perturbative components $\Phi^{(0)}$,
$\Phi^{(1)}$, and $\Phi^{(2)}$ of the LIT state.
This is achieved by systematically expanding all quantities in
Eq.~\eqref{eq:LIT-LS}, which includes the perturbative expansion of the
charge-density operator $\rho$:
\begin{equation}
 \rho = \rho^{(0)} + \rho^{(1)} + \rho^{(2)} + \cdots \,.
\end{equation}
Along with the expansion of the free Green's function (which follows directly
from the expansion of the initial energy),
\begin{equation}
 G_{0} = G_{0} - E_{i}^{(1)}G_{0}^{2}
 - \big[E_{i}^{(1)}\big]^2 G_0^{3} - E_{i}^{(2)} G_0^{2}
 + \cdots \,,
\label{eq:G0-series}
\end{equation}
we obtain altogether:
\begin{widetext}
\begin{subalign}
\label{eq:LIT-final-LO}
 \big(G_0 V^{(0)} - 1\big) \ket{\Phi^{(0)}}
 &= G_0 \rho^{(0)} \ket{\Psi_{i}^{(0)}} \,, \\
\label{eq:LIT-final-NLO}
 \big(G_0 V^{(0)} - 1\big) \ket{\Phi^{(1)}}
 &= G_0 \big(E_{i}^{(1)} - V^{(1)}\big) \ket{\Phi^{(0)}}
  + G_0 \Big[\rho^{(1)} \ket{\Psi_{i}^{(0)}}
  + \rho^{(0)} \ket{\Psi_{i}^{(1)}}\Big]  \,, \\
\label{eq:LIT-final-N2LO}
 \big(G_0 V^{(0)} - 1\big) \ket{\Phi^{(2)}}
 &= G_0 \big(E_{i}^{(2)} - V^{(2)}\big) \ket{\Phi^{(0)}}
  + G_0 \big(E_{i}^{(1)} - V^{(1)}\big) \ket{\Phi^{(1)}} \nonumber \\
 & \qquad \qquad \quad \quad \null
  + G_0 \Big[\rho^{(2)} \ket{\Psi_{i}^{(0)}}
  + \rho^{(0)} \ket{\Psi_{i}^{(2)}}
  + \rho^{(1)} \ket{\Psi_{i}^{(1)}} \Big] \,.
\end{subalign}
\end{widetext}
Written in this way, we can see that at each order in the perturbative
expansion, the LIT equations share the same ``kernel operator''
($G_0 V^{(0)} - 1$) and only differ in their inhomogeneous terms.
These inhomogeneous terms involve the lower-order LIT amplitudes, so ultimately
we have a scheme that can be evaluated sequentially.

\subsection{Chiral interaction}
\label{sec:Chiral}

We follow the power counting developed in Refs.~\cite{Long:2011xw, Long:2012ve,
Wu:2018lai} for the chiral potentials, and in Ref.~\cite{Shi:2022blm} for the
electric charge operators.
For completeness, we explain briefly here how these nuclear forces are arranged.

Generally, the potential derived from ChEFT at a given order receives
contributions from pion exchanges between the nucleons, as well as from contact
interactions that parameterize unresolved short-distance physics.
The expressions for pion-exchange potentials have become standardized and we use
Ref.~\cite{Epelbaum:1999dj} for the one-pion exchange (OPE) and the
leading two-pion exchange (TPE) potentials, \ie, Feynman diagrams made up of
vertices with chiral index $\nu = 0$, referred to as ``TPE0'' in
Ref.~\cite{Epelbaum:1999dj}.
One-pion exchange provides the longest-range component of the nuclear
force.\footnote{Electromagnetic forces behave like a power law at large
distances and therefore have a strictly longer range than OPE.
It is customary to consider these separate from the ``strong'' nuclear force,
although in fact the two cannot be strictly separated, see \eg
Ref.~\cite{Kong:1999sf}.
In any case, we do not need to consider electromagnetic forces between the
neutron and the proton in this work to the order we are working at.}
The values of the various parameters adopted here include the nucleon
axial coupling constant $g_A = 1.29$, the pion decay constant $f_\pi = 92.4$
MeV, the pion mass $m_\pi = 138$ MeV, and the average nucleon mass $m_N = 939$
MeV.

By definition, all LO forces are nonperturbative, \ie, they need to be fully
iterated by solving the Schr{\"o}dinger or Lippmann-Schwinger equations.
In line with this, the LIT expansion discussed in the previous section treats
$V^{(0)}$ nonperturbatively.

The OPE potential contributes to all $NN$ partial waves.
While it can be strongly attractive in some of these channels, as
angular momentum $l$ increases between the two nucleons the OPE potential is
gradually weakened by the centrifugal barrier and therefore becomes perturbative.
Following Refs.~\cite{Wu:2018lai, Kaplan:2019znu}, we choose \nnch1S0,
\nnch3S1-\nnch3D1, and \nnch3P0 to be the channels where OPE must be at LO,
and in each of these channels there is a contact potential, also referred to as
a counterterm in the literature, at this order.
For the $S$ waves, these contact potentials would anyway enter at LO, while for
\nnch3P0 NDA alone would suggest that the $P$-wave contact only enters as a
higher-order correction.
For each channel where OPE is attractive and singular, the Schr\"odinger equation
does not have a well-defined solution without an ultra-violet regularization,
even though the OPE potential itself is well-defined~\cite{Frank:1971xx}.
A counterterm needs to be included in each of these channels if one chooses
the potential to enter at
LO~\cite{Nogga:2005hy}.
Each of the contact terms discussed below are determined at the order in which
they appear via fitting to the empirical phase shifts for the corresponding
partial wave, provided by the Nijmegen partial-wave analysis~\cite{NNonline,
Stoks:1993tb}, up to $k = 300$ MeV where $k$ is the center-of-mass momentum.

In the \nnch1S0 and \nnch3S1 channels, the short-range, contact potentials $V_S$
at LO have the form
\begin{equation}
 \braket{p'; \alpha | V_S^{(0)} | p; \alpha} = C_{0}^{(0)} \,
\end{equation}
where $\alpha$ represents a given partial wave and $p$ ($p'$) is
the incoming (outgoing) relative momentum, as introduced in
Sec.~\ref{sec:PartialWaves}.
We use the ``$(0)$'' superscript to indicate LO potentials and LECs, as done for
other quantities in Sec.~\ref{sec:PertLIT}.
As the contact potentials are formulated directly in the partial-wave basis,
we omit the channel subscript $\alpha$ on the LECs when there is no
ambiguity.
In \nnch3P0, the contact potential features an explicit momentum dependence in
accordance with a generic $P$-wave short-range interaction:
\begin{equation}
 \braket{p';\nnch3P0 | V_S^{(0)} | p;\nnch3P0} = C_0^{(0)} p' p \,.
\end{equation}
In all other channels, the LO potential vanishes in the power counting of
Refs.~\cite{Wu:2018lai}.

At next-to-leading order (NLO), OPE enters in the channels where it is
considered perturbative.
Following further Refs.~\cite{Long:2011xw, Long:2012ve, Wu:2018lai}, we include
OPE in all partial waves with orbital angular momentum $l \leqslant 2$,
and those coupled to them, $\nnch3F2$ and $\nnch3G3$, deeming OPE in yet higher
partial waves to be negligible.
The cutoff on $l$ is chosen rather empirically, based on the observation that
the $F$-wave phases shifts in the partial-wave analysis, \eg, of
Ref.~\cite{Stoks:1993tb} are mostly $\lesssim 2^\circ$.
We note that this choice is in line with Ref.~\cite{PavonValderrama:2016lqn},
where OPE is counted as next-to-next-to-next-to-leading order (N3LO) in the
spin-singlet channels, and we stop at next-to-next-to-leading order (N2LO)
in this paper.
Unfortunately, similar analyses on the triplet channels are not yet available.
In addition, the \nnch1S0 contact potential receives a momentum-dependent
correction at NLO~\cite{Long:2012ve}:
\begin{equation}
 \braket{p';\nnch1S0 | V_S^{(1)} | p;\nnch1S0}
 = C_0^{(1)} + \frac{D_0^{(1)}}{2} ({p'}^2 + p^2) \,,
\end{equation}
where $C_0^{(1)}$ is the NLO correction to the LO LEC $C_0^{(0)}$
and $D_0^{(1)}$ is the LEC associated with the momentum-dependent \nnch1S0
contact interaction (which would naively enter at N2LO).

At N2LO, TPE0 enters in the channels where OPE enters at LO (\ie,
\nnch1S0, \nnch3S1--\nnch3D1, and \nnch3P0), while in other channels two-pion
exchange enters at higher order.
For the short-range potentials in these channels, the power counting gives the
following contributions at N2LO:
\begin{subalign}
 \braket{p';\nnch1S0 | V_S^{(2)} | p;\nnch1S0}
  &= E_0^{(2)} \, {p'}^2 p^2 \,, \\
 \braket{p';\nnch3S1 | V_S^{(2)} | p;\nnch3S1}
  &= C_0^{(2)} + \frac{D_0^{(2)}}{2} ({p'}^2 + p^2) \,, \\
 \braket{p';\nnch3D1 | V_S^{(2)} | p;\nnch3S1}
  &= E_0^{(2)} \, {p'}^2 \,.
\end{subalign}
In the channels where OPE is perturbative, we have the following $P$-wave
contact potentials at N2LO:
\begin{subalign}
 \braket{p';\nnch1P1 | V_S^{(2)} | p;\nnch1P1} &= C_0^{(2)} \, p' p \,, \\
 \braket{p';\nnch3P1 | V_S^{(2)} | p;\nnch3P1} &= C_0^{(2)} \, p' p \,, \\
 \braket{p';\nnch3P2 | V_S^{(2)} | p;\nnch3P2} &= C_0^{(2)} \, p' p \, .
\end{subalign}
In practical calculations, all interactions are regularized using a
momentum cutoff $\Lambda$, with $\Lambda\to\infty$ corresponding to the
zero-range limit.
We implement this cutoff regularization in a separable manner, \ie, we have,
generically,
\begin{equation}
 \braket{\alpha \, p' | V_\Lambda | \alpha p} \sim g(p') \, [\cdots] \, g(p) \,,
\end{equation}
where $[\cdots]$ represents the zero-range expression for the potential and
$g(p) = \exp({-}p^4/\Lambda^4)$.

\subsection{Current operator}
\label{sec:CurrentOp}

\subsubsection{Overview}

At LO, the charge density is given by a simple one-body operator
that describes a photon coupling to the nucleon charge:
\begin{multline}
 \rho^{(0)}(\vec{q};\vec{p},\vec{p'})
 = \braket{\vec{p}|\rho^{(0)}(\vecq)|\vec{p}'}
 = \ec\frac{1+\tau_3^{(1)}}{2} \\
 \null\times\vdelta\!\left(\vec{p}-\vec{p}'-\frac{\vecq}{2}\right)
 + 1 \rightleftharpoons 2 \, ,
 \label{eq:rho-0}
\end{multline}
There is no NLO correction to the charge-density operator, but at N2LO,
two corrections enter.
Following Refs.~\cite{Phillips:2003jz, Pastore:2011ip, Shi:2022blm}, we write
these as
\begin{subequations}
\begin{multline}
 \label{eq:rho-2-rel}
 \rho^{(2)}_{\text{rel}}(\vec{q};\vec{p},\vec{p'})
 = {-}\frac{\ec}{8\MN^2}\left(\frac{1}{2}+\kappa_s
 + (\frac{1}{2}+\kappa_v)\tau_3^{(1)}\right) \\
 \null\times \left(
  q^2+2\ii\vec{q}\cdot\vsigma_1\times\vec{K}
 \right)
 \vdelta\!\left(\vec{p}-\vec{p}'-\frac{\vecq}{2}\right)
 + 1 \rightleftharpoons 2
\end{multline}
and
\begin{equation}
 \label{eq:rho-2-str}
 \rho^{(2)}_{\text{str}}(\vec{q};\vec{p},\vec{p'})
 = {-}\frac{\ec}{6}\langle r^2_s\rangle q^2
 \vdelta\!\left(\vec{p}-\vec{p}'-\frac{\vecq}{2}\right) \,,
\end{equation}
\end{subequations}
where $\kappa_s (\kappa_v)$ and $r_s$ are, respectively, the isoscalar
(isovector) anomalous magnetic moment and the isoscalar charge radius of the
nucleon.
We have written Eqs.~\eqref{eq:rho-0} and~\eqref{eq:rho-2-rel} in terms of
nucleon ``1'' being struck by the photon, with ``$1 \rightleftharpoons 2$''
indicating that there is an analogous term to be added with superscript ``(2)''
on the isospin operator $\tau_3$; for the deuteron system that we consider here,
both terms ultimately give the same result and we account for this with a factor
two in our implementation.
Moreover, $\vec{K} = \vec{p}/2 + \vec{p'}/2 - \vec{q}/4$ in the c.m.\ frame of
the outgoing nucleons that we are working in.

\subsubsection{Relativistic considerations}
\label{sec:Relativity}

All the operators described above are defined in the two-nucleon c.m.\ frame,
and the expressions we have given follow from specializing the generic
expressions of Ref.~\cite{Pastore:2011ip} to this frame by inserting the
appropriate expressions for the kinematic variables.
One thing to note in this regard is that the resulting expressions for the
operators turn out to be formally the same as those for the form-factor
calculations in Ref.~\cite{Shi:2022blm}, even though that analysis is performed
in the so-called ``Breit frame,'' defined by the virtual photon transferring
only momentum and no energy to the deuteron.
One notable difference is  that in the Breit frame, $\vec{K} = \vec{p} +
\vec{q}/4$.
Seemingly, this would change the expression for
$\rho^{(2)}_{\text{rel}}(\vec{q};\vec{p},\vec{p'})$ in Eq.~\eqref{eq:rho-2-rel}.
However, by using the identity
\begin{equation}
 \vec{q}\cdot\vsigma_1\times\vec{K}
 = \vsigma_1\cdot\vec{K}\times\vec{q} \,,
\end{equation}
it becomes evident that any terms proportional to $\vec{q}$ in $\vec{K}$ drop
out, and then by means of overall momentum conservation one can easily transform
the Breit-frame expression into the c.m.-frame one.

While the three-momentum transfer $\vec{q}$ is in principle a frame-dependent
quantity, it is important to note that $q=\abs{\vecq}$ can be consistently (\ie,
up to differences that are of higher order than we work at in the EFT power
counting) assumed to be the same in both frames referred to above.
Accordingly, the discussion above is already exploiting this observation in simply referring
to the momentum transfer as $\vec{q}$, without indication which frame it should
refer to.
The argument for this equivalence goes as follows:
\begin{enumerate}
 \item In the c.m.\ frame, the outgoing neutron-proton pair is at rest, whereas
  prior to the reaction the deuteron moves with momentum ${-}\vec{q}$.
  In the Breit frame, on the other hand, the deuteron moves with momentum
  ${-}\vec{q}/2$ before being struck by the photon, and with momentum
  $\vec{q}/2$ afterwards.
  The relative motion between the two frames is pointing along the $z$-axis and
  can be written as $\beta = \Delta v/c = (v_{\text{c.m.}}-v_{\text{Breit}})/c$,
  where $c=1$ in our conventions.
  Then, with $\vec{q} = \Md\vec{v}$ (where $\Md = 2\MN-\Bd$ the deuteron mass),
  we can write
  \begin{equation}
   \beta = \frac{\Delta q}{\Md} = \OO(1/\MN) \,.
  \end{equation}
 \item A boost from the Breit frame to the c.m.\ frame simply amounts to
  $q \rightarrow \gamma q$ with $\gamma \approx 1+\beta^2/2$.
  In principle  there is another term ${-}\beta\gamma\omega$, but this vanishes
  identically if we start in the Breit frame, and it is otherwise suppressed by
  another factor $\beta \sim 1/\MN$.
  Using $\beta = \OO(1/\MN)$ as determined above, we find that the net change
  is of order $1/\MN^2$, \ie, it is of the same order as the explicit
  relativistic correction to the current, Eq.~\eqref{eq:rho-2-rel},
  which we include specialized to the c.m.\ frame.
 \item Considering that $Q^2 = {-}q_{\mu}q^{\mu} = \vecq^2 - \omega^2$ is the
  Lorentz-invariant squared four-momentum transfer, and $\omega$ being an
  explicit parameter in the c.m.\ frame (that in fact we use to define the
  response function and LIT), it might seem odd to write the contribution
  $\rho^{(2)}_{\text{str}}$ given in Eq.~\eqref{eq:rho-2-str} in terms of $q^2$
  only, when generally it should feature $Q^2$~\cite{Phillips:2003jz}.
  However, by the above analysis it is indeed consistent to formally
  count $\omega \sim \vecq^2/\MN$, and to consequently take
  $Q^2 = q^2 = \vec{q}^2$ as the same in all frames---up to corrections
  of order $1/\MN^2$ (see also Ref.~\cite{Krebs:2019aka} for this conclusion).
\end{enumerate}

\subsubsection{Boost correction}
\label{sec:CurrentOp-Boost}

One further correction to consider arises from the fact that we
perform the LIT calculation in the c.m.\ frame of the outgoing nucleons,
which, as already mentioned, before the collision is the frame that moves
with total momentum $-\vec{q}$, while calculation of the bound-state wave
function we use in this work (and that is standard) assumes the deuteron is
at rest (\ie, effectively, in the laboratory frame).
Consistency requires that this wave function before the collision must be
boosted into the c.m. frame, and, according to the EFT power counting
discussed below, this boost correction enters at N2LO.

The boost correction has already been studied in detail for calculations of the
deuteron form factors~\cite{Phillips:2003jz,Schiavilla:2002fq,Shi:2022blm}.
In that case, the situation is slightly different compared to the breakup
process because, as already mentioned in the previous section, form
factors are typically calculated in the Breit frame.
In the form-factor matrix element $\braket{\Psi|\mathcal{O}|\Psi}$,
where $\mathcal{O}$ denotes a generic current operator, the ket-side deuteron
then moves with momentum ${-}\vec{q}/2$ whereas the bra-side deuteron moves with
momentum ${+}\vec{q}/2$.
In this situation, boosts are required for the deuteron states on both sides if
their wave functions are calculated in the rest frame.

While in general the Lorentz boost operator acting on a deuteron state is quite
complicated (due to the deuteron being a spin-1 particle), it has been shown in
Ref.~\cite{Schiavilla:2002fq} that for the form-factor calculation the boost
effect can be reduced to merely a relativistic rescaling of the longitudinal
argument of the deuteron wave function expressed in momentum space (see
Ref.~\cite{Shi:2022blm} for a recent application of this).
The key to being able to perform this simplification is that spin-1 states
appear on both sides of the form-factor matrix element, while the relevant
current operators are odd functions under spin
exchange~\cite{Schiavilla:2002fq}.
For the breakup calculation we consider here, this remains true for the leading
current operator $\rho^{(0)}$, which we shall see is the only current operator
that enters together with the boost up to N2LO in the power counting.
Specifically, if we look at the explicit expression for the longitudinal
response function, Eq.~\eqref{eq:response-L}, we have matrix elements of the
form $\braket{\Psi_{f}|\rho^{(0)}|\Psi_{i}}$ in which $\ket{\Psi_{i}}$ needs to
be boosted.
Both $\ket{\Psi_i}$ and $\ket{\Psi_f}$ are $S=1$ states because
$\rho^{(0)}$ does not flip a nucleon spin, and therefore the situation is the
same as for the form-factor calculation.

The calculation of the boost correction as outlined above leads to expressions
of the form
\begin{equation}
 \braket{\vec{p}|\rho^{(0)}|\Psi_{i,\eta}}
 = \Psi_i\!\left(\vec{p}+\frac{\vecq}{2\sqrt{1+\eta}}\right) \,,
\label{eq:Psi-i-boosted}
\end{equation}
where $\eta=q^2/(4 M_N^2)$ describes the relativistic boost from the deuteron
rest frame to the c.m.\ frame (cf.\ the discussion in Sec.~\ref{sec:Relativity}),
and $\ket{\Psi_{i,\eta}}$ is the boosted deuteron state.
We have explicitly used here that the LO current $\rho^{(0)}$ amounts to merely
evaluating at a shifted momentum, and only this shift becomes modified by the
relativistic boost per the discussion above.
This ultimately allows us to model the boost correction, although strictly it
arises from the deuteron wave function, as an \emph{effective current
correction}.
That is, we can expand Eq.~\eqref{eq:Psi-i-boosted} to first order in
$\eta \sim Q^2/\MN^2$ to isolate the N2LO effect, and then construct an N2LO
operator $\rho^{(2)}_{\text{boost}}$ that, when applied to $\ket{\Psi_i}$,
reproduces the same expression.

One can obtain an expression for this operator from the aforementioned expansion
of the charge density into multipoles in a partial-wave momentum basis.
As detailed in Appendix~\ref{sec:CurrentPW-LO}, the (spin-decoupled) spatial
part of the matrix element of $\rho^{(0)}$ can be written in the form
\begin{multline}
 \braket{u;lm|\tilde{\rho}^{(0)}(\vecq)|u';l'm'}
 = \sum_{L} \sqrt{\hat{L}} \, C_{l m,L 0}^{l'm'} \\
 \null \times \int_{-1}^1 \dd x \, G^{L}_{l,l'}(u,q,x)
 \frac{\delta\big(u'-\iota(u,q,x)\big)}{u'^2} \,,
\label{eq:rho-0-G}
\end{multline}
where the function $G^{L}_{l,l'}(u,q,x)$ is defined in Eq.~\eqref{eq:G-rho-0}
and $\iota(u,q,x) = \sqrt{p^2 - p q x + q^2/4}$, as defined also in
Eq.~\eqref{eq:iota}.
To isolate the N2LO boost correction, we need to apply the substitution
\begin{equation}
 q \to \frac{q}{\sqrt{1+\eta}}
\end{equation}
in Eq.~\eqref{eq:rho-0-G} and then Taylor-expand the resulting expression to
first order.
We see from this there arises two distinct
terms: one from the $q$ dependence of $G^{L}_{l,l'}(u,q,x)$ and another from the
$\iota(u,q,x)$ inside the Dirac delta function.
The latter gives a term proportional to $\delta'\big(u'-\iota(u,q,x)\big)$,
which, when the operator is applied to a state, ultimately leads to a
first-order derivative of the momentum-space deuteron wave function.
Overall, the detailed expressions for the matrix elements of the effective
current operator $\rho^{(2)}_{\text{boost}}(\vecq)$ are given in
Appendix~\ref{sec:CurrentPW-Boost}.
To conclude this discussion, we note that the formalism described here can also
describe the boost required for the form-factor calculation, one merely needs to
make the replacement $\eta \to \eta_{\text{Breit}} = q^2/(4 M_d^2)$ to arrive at
the expression given in
Refs.~\cite{Phillips:2003jz,Schiavilla:2002fq,Shi:2022blm}.

\section{Results}
\label{sec:Results}

\subsection{Sanity checks}
\label{sec:FF-Sanity}

As explained in Sec.~\ref{sec:TheLIT}, for $L=0$ the second term in
Eq.~\eqref{eq:lit_der} is proportional to the square of the deuteron elastic
charge form factor, which can be calculated as
\begin{equation}
 F_C(q^2) = {\braket{\Psi ||\rho_0 || \Psi}} \,.
\label{eq:FC}
\end{equation}
Using $H\ket{\Psi} = E_{i}\ket{\Psi}$ along with
\begin{equation}
 \ket{\Phi_0} = (H - E_{i} - \sigma_R + i\sigma_I)^{-1} \rho_{0} \ket{\Psi}
\end{equation}
yields
\begin{equation}
 \braket{\Psi|\Phi_{0}}
 = \frac{\braket{\Psi| \rho_0 | \Psi}}{-\sigma_R + \ii\sigma_I} \,,
\end{equation}
which is what leads to Eq.~\eqref{eq:full-transform}.

This relation provides an excellent opportunity to check, numerically, the LIT
states $\ket{\Phi}$ against a direct calculation of $F_C(q^2)$.
Moreover, this comparison against a direct evaluation of the form factor can be
performed perturbatively at each order.
Obtaining a consistent form factor from ${\braket{\Psi|\Phi}}$ is a
necessary (although not sufficient) criterion for $\ket{\Phi}$ to be calculated
correctly.
Making use of our notation from Sec.~\ref{sec:PertLIT}, this leads to
\begin{equation}
 F_C^{(n)} = ({-}\sigma_{R}+\sigma_{I}) \BLIT^{(n)}
\end{equation}
for all orders $n$.

However, as discussed in Sec.~\ref{sec:CurrentOp}, at N2LO ($n=2$) one needs to
take into account relativistic corrections, and carefully consider the frame
in which the expressions are defined.
Per the discussion in Sec.~\ref{sec:Relativity}, all genuine current operators
at this order are formally the same in the Breit frame (the standard choice for
calculating the form factor) and the c.m.\ frame (used to calculate the LIT of
the longitudinal response function that describes the breakup process).
However, the boost correction detailed in Sec.~\ref{sec:CurrentOp-Boost}
\emph{does} depend on the frame, and therefore the direct equivalence breaks at
this point.

As we are merely looking to check our numerical implementation, we perform the
following comparison with the boost correction disabled at N2LO.
We have independently verified that our direct calculation of the form factor,
using Eq.~\eqref{eq:FC}, perturbatively expanded, agrees well with the
calculation of \textcite{Shi:2022blm}.
There is no NLO column in the table because there is no NLO correction to the
form factor: such a correction would be generated only by an NLO contribution to
the \nnch3S1-\nnch3D1 potential or by an NLO current operator, both of which are
absent.
For the LIT calculations (shown in the following section), there \emph{are} NLO
corrections that stem from the NLO \nnch1S0 potential contribution.

\begin{table*}
\centering
\pgfplotstabletypeset
[
  columns/q2/.style={column type=c, column name=$q^2$ ($m_{\pi}$)},
  columns/lit0/.style={column type=|c, column name=LIT (LO),
  fixed, zerofill, precision=5},
  columns/drt0/.style={column type=c, column name=direct (LO),
  fixed, zerofill, precision=5},
  columns/lit2/.style={column type=|c, column name=LIT (N2LO$^*$),
  fixed, zerofill, precision=5},
  columns/drt2/.style={column type=c, column name=direct (N2LO$^*$),
  fixed, zerofill, precision=5},
  every head row/.style={before row=\toprule,after row=\midrule},
  every last row/.style={after row=\bottomrule}
]
{
  q2  lit0      drt0        lit2      drt2
  0.5 0.928878  0.928877  -0.0101825 -0.010183
  1.0 0.768096  0.768093  -0.0339437 -0.033944
  2.0 0.595886  0.595882  -0.0597546 -0.059755
  3.0 0.447855  0.447851  -0.0804295 -0.080430
  4.0 0.239351  0.239348  -0.0977366 -0.097740
  5.0 0.117475  0.117472  -0.0858984 -0.085914
  6.0 0.0483624 0.048360  -0.055781  -0.055804
}
\caption{
Form factors extracted from the LIT calculation compared to a direct evaluation.
As discussed in the text, the boost correction that enters at N2LO differs
between the LIT and a direct form-factor calculation.
The values in the table therefore omit the boost correction at N2LO, indicated
with an asterisk.
\label{tab:FC-Checks}
}
\end{table*}

\subsection{Convergence studies}
\label{sec:Convergence}

In this section, we study how the LIT of the deuteron longitudinal response
function converges with respect to the order of the chiral expansion, and at
each order we also verify that the calculation converges as the momentum cutoff
$\Lambda$ increases, indicating that the theory is properly renormalized.
The Chiral EFT interaction that we are using has been shown to describe well
both two-nucleon scattering and deuteron form factors, and to converge for these
observables in the sense mentioned above, but for the electrodisintegration
process we are studying here, this still needs to be tested explicitly.

\begin{figure}[tbhp]
 \centering
 \includegraphics[width=1.0\columnwidth]{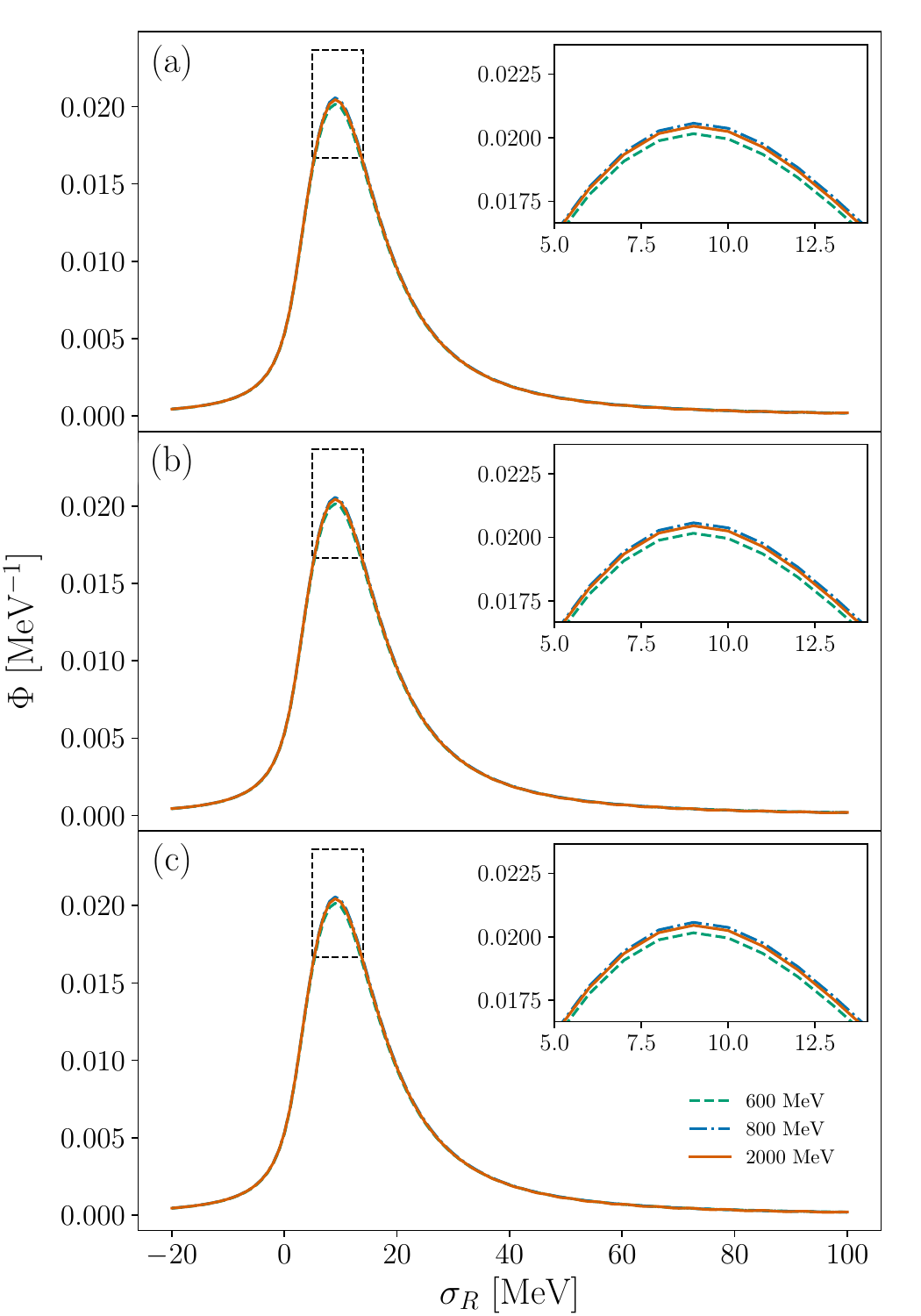}
 \caption{%
  LIT of the deuteron longitudinal response for $\vecq^2 = 0.6~\fm^{-2}$,
  $\sigma_I = 3.0~\MeV$ at (a) LO (b) NLO and (c) N2LO.
  Each panel shows the LIT for three different values of the momentum
  cutoff $\Lambda$.
  \label{fig:cut-conv}
 }
\end{figure}

\begin{figure}[tbhp]
 \centering
 \includegraphics[width=0.95\columnwidth]{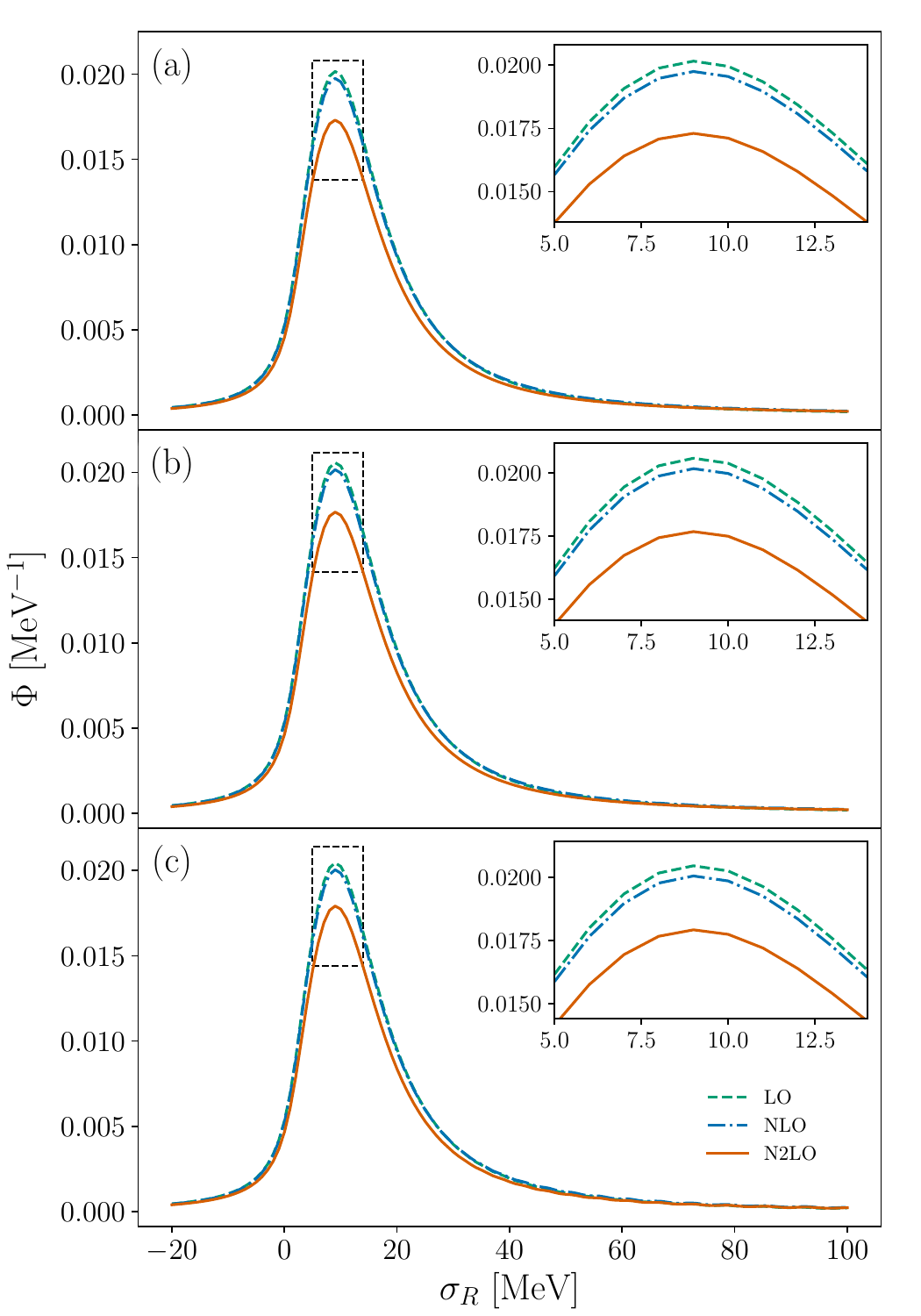}
 \caption{%
  LIT of the deuteron longitudinal response at different orders for $\sigma_I =
  3.0~\MeV$, $\vecq^2 = 0.5~\fm^{-2}$.
  \label{fig:order-cutoff}
 }
\end{figure}

In Fig.~\ref{fig:cut-conv} we show a comparison of the LIT calculated at
different EFT orders, including corrections to both the potential as well as to
the current operator (as discussed in Sec.~\ref{sec:Chiral}).
At each order, we show results for several cutoffs $\Lambda$.
For these calculations, which use $\vecq^2 = 0.6~\MeV^2$ for the momentum
transfer and set $\sigma_I = 3.0$MeV for the LIT.
We find good convergence with the cutoff, with very little variation
beyond $\Lambda = 600~\MeV$.
In Fig.~\ref{fig:order-cutoff}, we show the same data as in
Fig.~\ref{fig:cut-conv} in a different form, comparing the different EFT orders
at a fixed EFT cutoff value $\Lambda = 800\,\text{MeV}$, just to further illustrate
our findings.


Convergence with respect to the order-by-order EFT expansion is a bit more
difficult to interpret, and we note that the LIT at NLO is very close to the LO
result.
The reason for this is that, in the power-counting scheme we are using, at NLO
there are corrections only to the potential, namely a short-range contact term
in the \nnch1S0 channel, as well as OPE in the channels where it did not enter
already at LO.
In particular, there are no potential corrections in the \nnch3S1--\nnch3D1
channel (and thus no corrections to the deuteron wave function) or to the
current (charge density) operator at this order.
For the LIT of the longitudinal response, these potential corrections \emph{do}
play a role, but the effect of this final-state interaction\footnote{%
In a direct calculation of the response function, it becomes more obvious that
there is a correction at NLO to only the ``final-state interaction'' between the
proton and neutron arising from the deuteron breaking up.}
is suppressed in the kinematic region of the quasi-elastic peak (see, for example,
Refs.~\cite{More:2015tpa,More:2017syr} for discussions of this), and therefore
not very pronounced in our plotted results.
If one studies the NLO correction relative to LO, then indeed as one moves away
from the peak (\eg, by increasing $\sigma_R$ at fixed $q^2$), one can see that
the NLO correction becomes more sizable.

At N2LO, there are corrections to the potential in various channels, and, as
discussed in Sec.~\ref{sec:Chiral}, also to the current operator.
Overall these effects combine to decrease the magnitude of the LIT, which is
particularly notable in the peak region.
At this order, there is a bit more spread as the cutoff is varied, but
overall we still observe rapid convergence with increasing $\Lambda$.

\begin{figure}[tbhp]
 \centering
 \includegraphics[width=0.95\columnwidth]{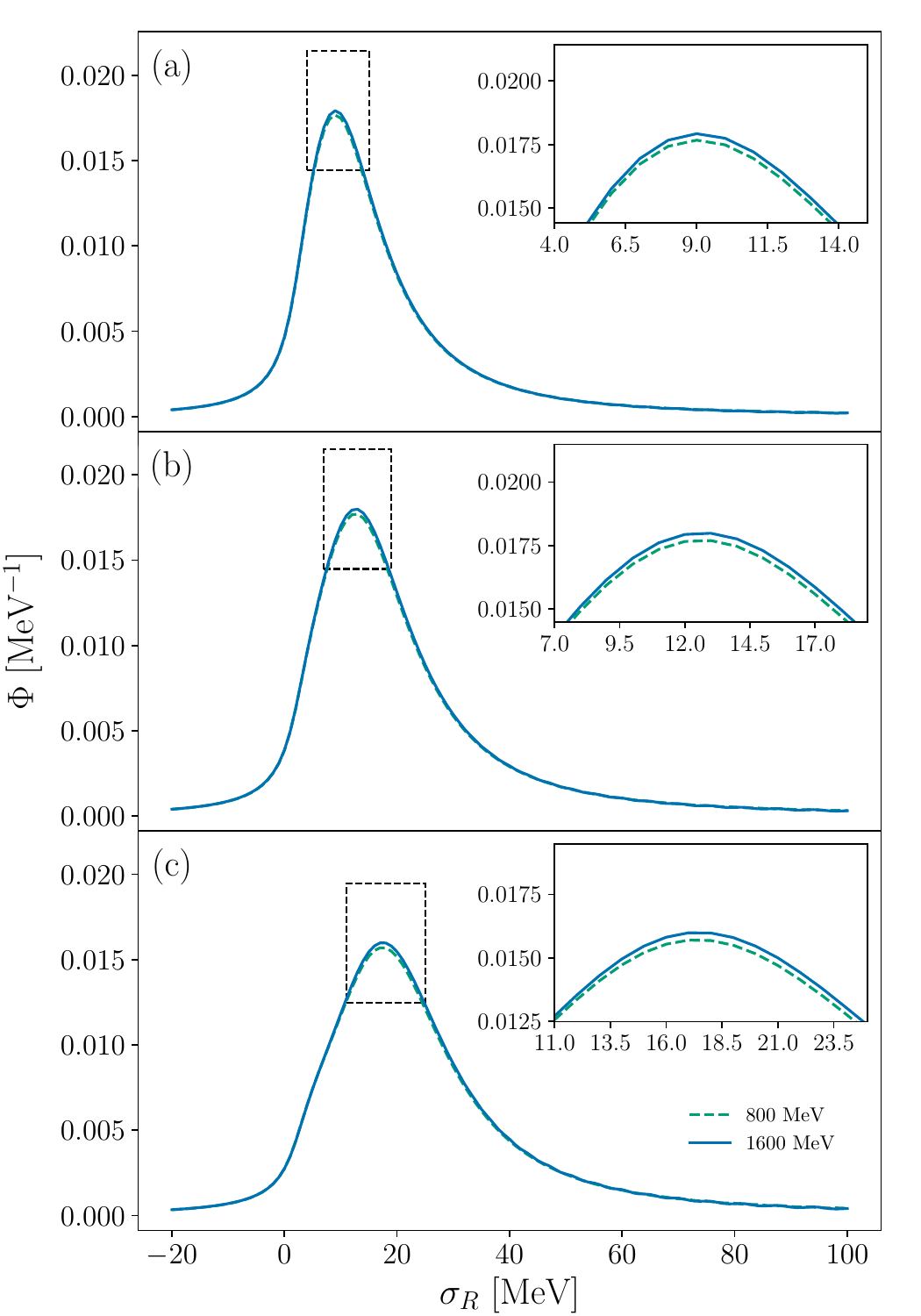}
 \caption{
  N2LO LIT of the deuteron response for $\Lambda = 1600$ MeV,
  $\sigma_I = 3.0$ MeV for momentum transfer (a) $\vecq^2 = 0.1$ fm$^{-2}$,
  (b) $\vecq^2 = 0.5$ fm$^{-2}$ and (c) $\vecq^2 = 1.0$ fm$^{-2}$.
  \label{fig:mom-trans}
 }
\end{figure}
\begin{figure}[tbhp]
 \centering
 \includegraphics[width=0.95\columnwidth]{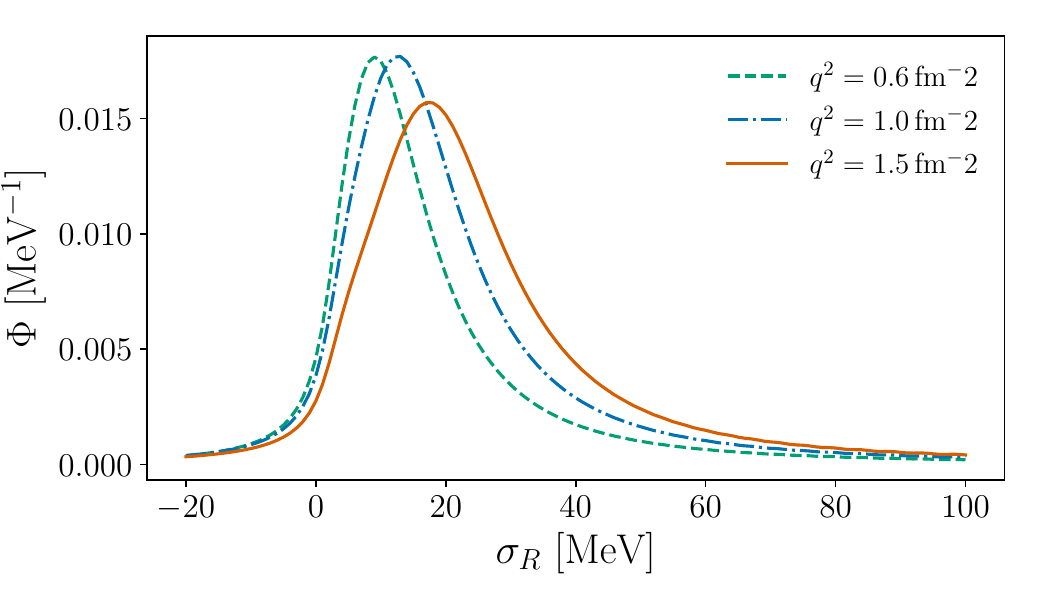}
 \caption{
  N2LO LIT of the deuteron response for $\Lambda = 800$ MeV,
  $\sigma_I = 3.0$ MeV for momentum transfer $\vecq^2 = 0.6$ fm$^{-2}, 1.0$
  fm$^{-2}, 1.5$ fm$^{-2}$.
  \label{fig:var-mom}
 }
\end{figure}

\subsection{Confronting experimental data}
\label{sec:ExpData}

Having analyzed in detail the convergence and systematics of our LIT
calculation, we can finally confront experimental data to assess our results.
In particular, this comparison provides new insights into the performance of the
perturbatively renormalized Chiral EFT interaction that we are using.
In order to carry out this comparison, we need to invert the LIT and from it
reconstruct the original response function.
There is ample discussion in the literature that explains how delicate this
inversion is in practice, as it falls into a category of technically ill-posed
problems~\cite{Efros:2007nq}.
The issue is that multiple different response functions can lead to all but
identical Lorentz transforms, so that as a mapping of functions the LIT is
almost not invertible -- especially in the practical numerical calculations,
where one always only ever has the functions defined on finite domains, and
where one generally has to work with finite-precision floating-point numbers.
A typical phenomenon is that the inversion can lead to high-frequency
oscillations in the reconstructed response, which get averaged out if one
completes the circle by applying again the LIT~\cite{Efros:2007nq}.

All that said, with sufficient care it \emph{is} possible to perform successful
LIT inversions.
We employ here a variant of the expansion method introduced already in
Ref.~\cite{Efros:1994iq}, the details of which are explained in
Appendix~\ref{sec:Inversion}.
In Fig.~\ref{fig:orders-data-0p6} we compare the response obtained from
inverting the LIT, calculated for momentum transfer $\vecq^2 = 0.6$ fm$^{-2}$
with EFT cutoff $\Lambda = 800~\MeV$ and $\sigma_I = 3.0~\MeV$ to experimental
data of \textcite{Simon:1979bu}.\footnote{Specifically, this data has been
take from Table 1 in Ref.~\cite{Simon:1979bu}.}
For this comparison we note that different conventions for expressing the
electrodisintegration cross section in terms of response functions exist in the
literature.
For the theory calculation, we followed
Refs.~\cite{Forest:1966,Carlson:1997qn,Bacca:2014,Martinelli:1995vn}, while
Ref.~\cite{Simon:1979bu} uses the convention of
\textcite{Fabian:1979kx}.\footnote{%
A comment in Ref.~\cite{Fabian:1979kx} furthermore suggests that the factor
$\omega^{c.m.}$ in Eq.~(1) of Ref.~\cite{Simon:1979bu} is a typographical
error, and consequently we dropped it in order to match conventions to ours.
}
A careful comparison of conventions (see also Ref.~\cite{Arenhovel:2004bc})
along with reducing all kinematic quantities non-relativistically (consistent
with our perturbative treatment of relativistic corrections) ultimately yields
that we need to multiply our response function by a factor $2\pi^2\alpha$ in
order to compare to the experimental data of \textcite{Simon:1979bu}.

\begin{figure}[tbhp]
 \centering
 \includegraphics[width=1.0\columnwidth]{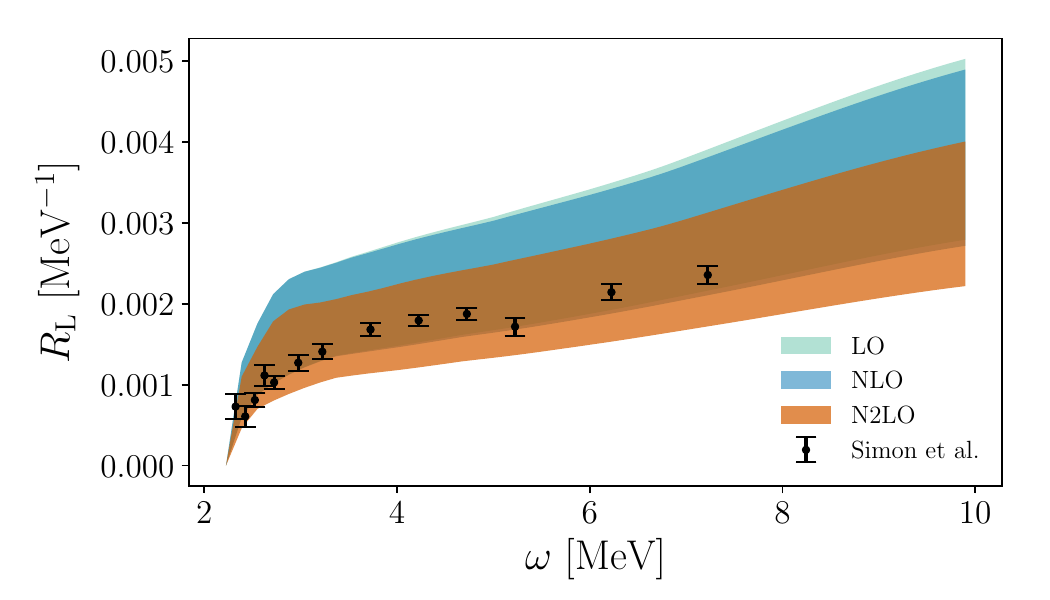}
 \caption{
  Deuteron response from the inverted LIT $\Lambda = 800$ MeV, $\sigma_I = 3.0$ MeV
  for momentum transfer $\vecq^2 \approx 1.0$ fm$^{-2}$ for orders LO, NLO, N2LO.
  The experimental data presented are from Ref.~\cite{Simon:1979bu}.
  \label{fig:orders-data-1p0}
 }
\end{figure}

Including this factor, we obtain the results shown in
Figs.~\ref{fig:orders-data-0p6} and ~\ref{fig:orders-data-1p0} for the
longitudinal response function at $q^2 = 0.6\,\text{fm}^{-2}$ and
$q^2 = 1.0\,\text{fm}^{-2}$, respectively.
In these figures, we include two sources of uncertainty: the uncertainty due
to the inversion procedure and the EFT uncertainty.
As outlined in Appendix~\ref{sec:Inversion}, the inversion procedure
generates a band with upper and lower bounds $R_{\text{max}/\text{min}}$.
To include the EFT uncertainty in addition, we apply factors $1\pm\lambda_n$
to the upper and lower bounds of the inversion bands, with
$\lambda_n = (Q/M_{\text{hi}})^{n+1}$ and $n$ the EFT order.
For the breakdown scale, we again follow Ref.~\cite{Shi:2022blm} and use
the delta-nucleon mass splitting, $M_{\text{hi}} = \delta \simeq 300\,\text{MeV}$,
as a conservative estimate.
For deuteron disintegration, we follow Ref.~\cite{Christlmeier:2008ye} and
identify the outgoing momentum of the $np$-pair,
\begin{equation}
 \abs{\vecp'} = \sqrt{(\omega - B)M_N + \vecq^2\frac{M_N}{2M_d}}
\end{equation}
as one relevant scale for the EFT analysis, in addition to the momentum
transfer $\abs{\vecq}$.
For estimating the uncertainty, we set
\begin{equation}
 Q = \text{max}(\abs{\vecp'}, \abs{\vecq})
\end{equation}
to determine $\lambda_n$.
Overall we find that our theoretical calculation of the response function
agrees well with the experimental data, and that it exhibits good convergence
with respect to the Chiral EFT expansion.

\begin{figure}[tbhp]
 \centering
 \includegraphics[width=1.0\columnwidth]{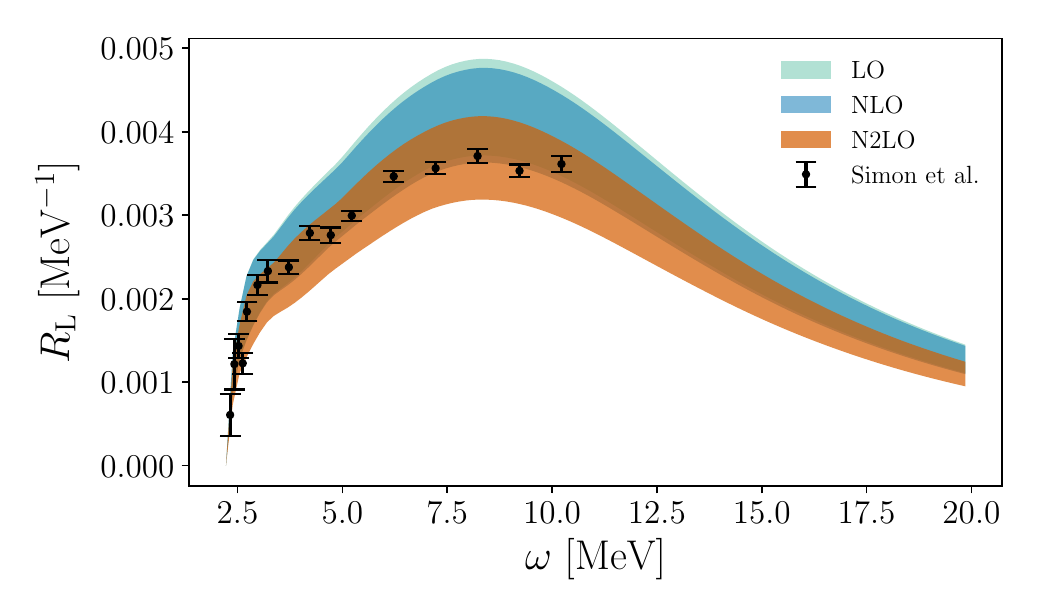}
 \caption{
  Deuteron response from the inverted LIT for $\Lambda = 800$ MeV, $\sigma_I = 3.0$ MeV
  for momentum transfer $\vecq^2 \approx 0.6$ fm$^{-2}$ for orders LO, NLO, N2LO.
  The experimental data presented are from Ref.~\cite{Simon:1979bu}.
  \label{fig:orders-data-0p6}
 }
\end{figure}

\section{Summary and outlook}
\label{sec:Conclusion}

In this work, we have studied deuteron electrodisintegration in Chiral EFT,
using the power counting of Refs.~\cite{Long:2011xw, Long:2012ve, Wu:2018lai,
Shi:2022blm} to achieve a consistently renormalized formulation of the theory.
To maintain this renormalization order by order, we follow a rigorously perturbative
approach that treats only LO non-perturbatively and includes all higher-order
corrections in distorted-wave perturbation theory.
As we utilize the LIT method in order to calculate the longitudinal response
function for the electrodisintegration process, a key result of our work is the
derivation of the perturbative expansion for the LIT equations, which we formulate
in momentum space, inspired by Ref.~\cite{Martinelli:1995vn}.
Our numerical calculations demonstrate that the methods work well, and, by studying
variation of the EFT cutoff, we find that the theory is properly renormalized for
the breakup observable that we study.
To the best of our knowledge, this has not been previously tested within this
framework.
Moreover, we find that our EFT results are in good agreement with experimental
data available within the range of energies that we would (Delta-less) Chiral EFT
to be valid in.
Our results therefore pave the way for future studies of inelastic processes in
a range of nuclei, using perturbatively renormalized Chiral EFT, either in the
``minimally modified Weinberg (MMW)'' approach adopted in this work, and also
using recently suggested novel schemes for Chiral EFT with fully perturbative
pions~\cite{Teng:2024exc, Lyu:2025yhz}.
Concretely, we plan concrete investigations along these lines for light nuclei
($A \geq 3$), which can be treated within the momentum-space
Faddeev/Faddeev-Yakubowsky framework of Refs.~\cite{Konig:2019xxk, Peng:2021pvo,
Lyu:2025yhz}.
In particular, it would be interesting to study the monopole transition in
\isotope[4]{He}, which has received a lot of attention in recent
years~\cite{Bacca:2012xv, Kegel:2021jrh, Michel:2023ley, Meissner:2023cvo,
Arenhovel:2004bc, Yin:2024xsg},
within our framework.

\begin{acknowledgments}
We thank Hang Yu, Nuwan Yapa, Winfried Leidemann, Giuseppina Orlandini,
Sonia Bacca, Francesca Bonaiti, Xincheng Lin, Atul Kedia, and Daniel Phillips for
useful discussions.
We furthermore thank Rui Peng for preparing the matrix elements of the chiral
potentials used in this work.
This work was supported in part by the U.S.\ National Science Foundation (Grant
No. PHY--2044632) and by the National Natural Science
Foundation of China (NSFC) under the Grants Nos. 12275185 and 12335002.
This material is based upon work supported by the U.S.\ Department of Energy,
Office of Science, Office of Nuclear Physics, under the FRIB Theory Alliance,
Award No.\ DE-SC0013617.
Computational resources for parts of this work were provided by the
high-performance computing cluster operated by North Carolina State University.
\end{acknowledgments}

\appendix

\section{Perturbative expansion of the deuteron wave function}
\label{sec:DeuteronWF}

As mentioned in the main text, evaluating the LIT in a perturbative
manner relies on a perturbative expansion of the deuteron bound state that
appears on the right-hand side of Eq.~\eqref{eq:LIT-LS}.
To find this perturbative expansion, one can start
from the Schr\"odinger equation written as
\begin{equation}
 (E - H_0) \ket{\psi} = V \ket{\psi} \,.
\label{eq:SEq}
\end{equation}
Expanding both $E = E^{(0)} + E^{(1)} + E^{(2)} + \cdots$ and
$\ket{\psi} = \ket{\psi^{(0)}} + \ket{\psi^{(1)}} + \ket{\psi^{(2)}} + \cdots$,
along with $V$ as described in Eq.~\eqref{eq:V-series} leads to equations of
the form
\begin{subalign}
 (E^{(0)} - H_0) \ket{\psi^{(0)}} &= V^{(0)} \ket{\psi^{(0)}} \\
 (E^{(0)} - H_0) \ket{\psi^{(1)}}
  &= V^{(0)} \ket{\psi^{(1)}} \nonumber \\
  &\qquad + (V^{(1)} - E^{(1)}) \ket{\psi^{(0)}}  \\
 (E^{(0)} - H_0) \ket{\psi^{(2)}}
  &= V^{(0)} \ket{\psi^{(2)}} \nonumber \\
  &\qquad + (V^{(1)} - E^{(1)}) \ket{\psi^{(1)}} \nonumber \\
  &\qquad + (V^{(2)} - E^{(2)}) \ket{\psi^{(0)}} \,.
\end{subalign}
Using the definition of the free Green's function,
\begin{spliteq}
  G_{0}(z) = (z - H_{0})^{-1} \,,
\end{spliteq}
evaluated at the leading order binding energy $z=E^{(0)}=-B^{(0)}$,
these equations can be rewritten as
\begin{subalign}
 (G_{0} V^{(0)} - 1) \ket{\psi^{(0)}} &= 0 \\
 (G_{0} V^{(0)} - 1) \ket{\psi^{(1)}}
  &= G_{0}\, (E^{(1)} - V^{(1)}) \ket{\psi^{(0)}}  \\
 (G_{0} V^{(0)} - 1) \ket{\psi^{(2)}}
  &= G_{0} \,(E^{(1)} - V^{(1)}) \ket{\psi^{(1)}} \nonumber \\
  &\qquad + G_{0} \,(E^{(2)} - V^{(2)}) \ket{\psi^{(0)}} \,.
\end{subalign}
Alternatively, one can directly rewrite Eq.~\eqref{eq:SEq} in terms of $G_0(E)$
and perturbatively expand the resulting expression.

\section{Multipole and partial-wave expansion of the current operators}
\label{sec:CurrentPW}

We wish to expand any generic operator $\rho(\vecq)$ into multipoles, as
stated in Eq.~\eqref{eq:rho-L} in the main text.
This expansion will arise naturally from expressing $\rho(\vecq)$ in the
momentum-space partial-wave basis, the details of which are provided in the
following subsections.

\subsection{Leading-order charge operator}
\label{sec:CurrentPW-LO}

For the LO charge density operator $\rho^{(0)}(\vecq)$, we start by separating
the spatial and discrete parts as
\begin{equation}
 \rho^{(0)}(\vecq) = \tilde{\rho}^{(0)}(\vecq) \sum_{i=1,2}
\frac{1+\tau_3^{(i)}}{2} \,,
\end{equation}
where $\tau_3^{(i)}$ is a Pauli matrix acting on nucleon $i$ in isospin space,
and
\begin{equation}
 \braket{\vecp|\tilde{\rho}^{(0)}(\vecq)|\vecp'}
 = \vdelta\big(\vecp - \vecp' - \tfrac12\vecq\big) \,,
\end{equation}
in momentum space, where the factor $1/2$ stems from embedding the one-body
operator into a two-nucleon space expressed in relative coordinates.

We use $u$ in the following to denote momenta and write two-nucleon partial-wave
states as $\ket{u;\alpha} = \ket{u;(ls)j,t}$.
To further evaluate the charge density operator, we need to decouple the spatial
part according to
\begin{equation}
 \ket{u;\couple{l}{s}{j};t}
 = \sum_{m,m_s} \CG*{lm}{sm_s}{j\cdot} \ket{u;lm}\ket{sm_s;t} \,.
 \label{eqn:statebs}
\end{equation}
Note that we are omitting here the projection quantum number associated with the
total angular momentum $j$, indicated with the `$\cdot$' on the Clebsch-Gordan
coefficient.

Focusing first on the spatial part of the operator, we have to evaluate
\begin{spliteq}
 &\braket{u;lm|\tilde{\rho}^{(0)}(\vecq)|u';l'm'} \\
 &= \int\!\dd^3p\int\!\dd^3p'
  \,\braket{u;l m|\vecp} \braket{\vecp|\hat{\rho}^{(0)}(\vecq)|\vecp'}
  \braket{\vecp'|u';l'm'} \\
 &= \int\!\dd^3p\int\!\dd^3p'
  \,\YY_{l m}(\hat{\vecp}) \frac{\delta(u-p)}{u^2}
  \vdelta\big(\vecp-\vecp'-\tfrac12\vecq\big) \\
 &\hspace{6em}\null\times\YY_{l'm'}^*(\hat{\vecp}')
  \frac{\delta(u'-p')}{u'^2} \\
 &= \int\!\dd p\,p^2\int\!\dd\Omega_{p}
  \,\YY_{lm}(\hat{\vecp}) \frac{\delta(u-p)}{u^2}
  \frac{\delta(u'-\abs{\vecp-\tfrac12\vecq})}{u'^2} \\
 &\hspace{6em}\null
  \times \YY_{l'm'}^*\big(\reallywidehat{\vecp-\tfrac12\vecq}\big) \,.
\label{eq:rho-ulm-0}
\end{spliteq}
Above the names of vectors, we use the hat symbol to denote the angular part in
a spherical representation of the vector.

We proceed to modify this expression, using techniques described in
Ref.~\cite{Gloeckle:1983} for the permutation operators appearing in the Faddeev
equations, and expanding upon a similar discussion in Ref.~\cite{Konig:2019xxk}
(which we reproduce here to keep this section self-contained) to keep track
of all multipoles.
To begin, we note that
\begin{multline}
 \frac{\delta(u'-\abs{\vecp-\tfrac12\vecq})}{u'^2}
 = 2\pi \sum_{k} \sqrt{\hat{k}} ({-}1)^k \Bigg[
  \int_{-1}^1 \dd x\,P_k(x) \\
  \null \times \frac{\delta\big(u'-\sqrt{p^2-pqx+q^2/4}\big)}{u'^2} \Bigg]
  \YYY_{kk}^{00}(\hat{\vecp},\hat{\vecq}) \,,
\end{multline}
where $\YYY_{l_1,l_2}^{LM}$ is used to denote two coupled
spherical harmonics.
Moreover,
\begin{multline}
 \YY_{l'm'}^*\big(\reallywidehat{\vecp-\tfrac12\vecq}\big)
 = \sum_{\lambda_1'+\lambda_2'=l'}
  \frac{p^{\lambda_1'}\big({-}\tfrac12q\big)^{\lambda_2'}}
  {\abs{\vecp-\tfrac12\vecq}^{l'}} \\
  \null \times \sqrt{\frac{4\pi(2l'+1)!}{(2\lambda_1'+1)!(2\lambda_2'+1)!}}
  \YYY_{\lambda_1'\lambda_2'}^{l'm'*}(\hat{\vecp},\hat{\vecq}) \,.
\end{multline}
Inserting these into Eq.~\eqref{eq:rho-ulm-0} gives
\begin{widetext}
\begin{multline}
 \braket{u;lm|\tilde{\rho}^{(0)}(\vecq)|u';l'm'}
 = \int\!\dd p p^2 \dd\hat{\vecp}
  \,\YY_{l m}(\hat{\vecp}) \frac{\delta(u-p)}{u^2} \\
 \null \times 2\pi \sum_{k} \sqrt{\hat{k}} ({-}1)^k \Bigg[
   \int_{-1}^1 \dd x\,P_k(x)
   \frac{\delta\big(u'-\sqrt{p^2-pqx+q^2/4}\big)}{u'^2} \Bigg]
   \YYY_{kk}^{00}(\hat{\vecp},\hat{\vecq}) \\
 \null \times \sum_{\lambda_1'+\lambda_2'=l'}
  \frac{p^{\lambda_1'}\big({-}\tfrac12q\big)^{\lambda_2'}}
  {\abs{\vecp-\tfrac12\vecq}^{l'}}
  \sqrt{\frac{4\pi(2l'+1)!}{(2\lambda_1'+1)!(2\lambda_2'+1)!}}
  \YYY_{\lambda_1'\lambda_2'}^{l'm'*}(\hat{\vecp},\hat{\vecq}) \,.
\label{eq:rho-ulm-1}
\end{multline}
\end{widetext}
The product of two coupled spherical harmonics can be reduced as follows:
\begin{multline}
 \YYY_{\lambda_1'\lambda_2'}^{l'm'*}(\hat{\vecp},\hat{\vecq})
 \YYY_{kk}^{00}(\hat{\vecp},\hat{\vecq})
 = \frac{1}{4\pi} \sqrt{\hat{k}\hat{\lambda_1'}\hat{\lambda_2'}}
  ({-}1)^{\lambda_1'+\lambda_2'+l'} \\
 \null \times \sum_{f_1,f_2}
  \SixJ{f_2}{f_1}{l'}{\lambda_1'}{\lambda_2'}{k}
  \CG*{k0}{\lambda_1'0}{f_10} \CG*{k0}{\lambda_2'0}{f_20}
  \YYY_{f_1f_2}^{l'm'*}(\hat{\vecp},\hat{\vecq}) \,.
  \label{eqn:conct_biH}
\end{multline}
We can now perform the integral over $\hat{\vecp}$ in
Eq.~\eqref{eq:rho-ulm-1}:
\begin{equation}
 \int\!\dd\Omega_{p} \YY_{l m}(\hat{\vecp})
 \YYY_{f_1f_2}^{l'm'*}(\hat{\vecp},\hat{\vecq}) \\
 = \sum_{m_2} C_{l m,f_2 m_2}^{l'm'} \YY_{f_2 m_2}^*(\hat{\vecq})
  \,\delta_{f_1l} \,.
  \label{eqn:int_p}
\end{equation}
Noting that choosing $\vecq$ aligned with the $\hat{z}$ axis collapses the sum
to just $m_2=0$ and then renaming $f_2 \to L$ establishes the connection with
Eq.~\eqref{eq:rho-L}, and we have
\begin{equation}
 \YY_{L0}^*(q\hat{z}) = \frac{\sqrt{\hat{L}}}{\sqrt{4\pi}} \,.
 \label{eqn:YY_qz}
\end{equation}
Overall, collecting the various terms, we arrive at
\begin{widetext}
\begin{multline}
 \braket{u;lm|\tilde{\rho}^{(0)}(\vecq)|u';l'm'}
 = \int\!\dd p\,p^2 \frac{\delta(u-p)}{u^2}
 \sum_{L} \sqrt{\hat{L}} \, C_{l m,L 0}^{l'm'}
 \null \times \frac12 \sum_{k} \sqrt{\hat{k}} ({-}1)^k \\
  \int_{-1}^1 \dd x\,P_k(x)
  \frac{\delta\big(u'-\sqrt{p^2-pqx+q^2/4}\big)}{u'^2}
 \sum_{\lambda_1'+\lambda_2'=l'}
  \frac{p^{\lambda_1'}\big({-}\tfrac12q\big)^{\lambda_2'}}
  {\abs{\vecp-\tfrac12\vecq}^{l'}}
  \sqrt{\frac{(2l'+1)!}{(2\lambda_1'+1)!(2\lambda_2'+1)!}}
 \sqrt{\hat{k}\hat{\lambda_1'}\hat{\lambda_2'}}
  ({-}1)^{\lambda_1'+\lambda_2'+l'} \\
 \null \times \SixJ{L}{l}{l'}{\lambda_1'}{\lambda_2'}{k}
  \CG*{k0}{\lambda_1'0}{l0} \CG*{k0}{\lambda_2'0}{L0} \,,
\label{eq:rho-ulm-L}
\end{multline}
which can be grouped as
\begin{equation}
 \braket{u;lm|\tilde{\rho}^{(0)}(\vecq)|u';l'm'}
 = \sum_{L} \sqrt{\hat{L}} \braket{u;lm|\tilde{\rho}^{(0)}_{L}(q)|u';l'm'}
 = \sum_{L} \sqrt{\hat{L}} \, C_{l m,L 0}^{l'm'}
 \braket{u;l||\tilde{\rho}^{(0)}_{L}(q)||u';l'} \,.
\end{equation}
\end{widetext}
This is exactly the form that is to be expected from the Wigner-Eckart theorem,
and we note that the Clebsch-Gordan coefficient $C_{l m,L 0}^{l'm'}$
implies $m = m'$, which is a consequence of our choice of $\vecq$ aligned with
the $z$ axis.
Eliminating also the integral over $p$, the final result for the reduced
matrix element can be written as
\begin{multline}
 \braket{u;l||\tilde{\rho}^{(0)}_{L}(q)||u';l'}
 = \int_{-1}^1 \dd x \, G^{L}_{l,l'}(u,q,x)
 \frac{\delta\big(u'-\iota(u,q,x)\big)}{u'^2}
\end{multline}
with
\begin{widetext}
\begin{multline}
 G^{L}_{l,l'}(u,q,x)
 = \frac12 \sum_{k} \sqrt{\hat{k}} ({-}1)^k P_k(x)
 \sum_{\lambda_1'+\lambda_2'=l'}
  \frac{u^{\lambda_1'}\big({-}\tfrac12q\big)^{\lambda_2'}}
  {\iota(u,q,x)^{l'}}
  \sqrt{\frac{(2l'+1)!}{(2\lambda_1'+1)!(2\lambda_2'+1)!}}
 \sqrt{\hat{k}\hat{\lambda_1'}\hat{\lambda_2'}}
  ({-}1)^{\lambda_1'+\lambda_2'+l'} \\
 \null \times \SixJ{L}{l}{l'}{\lambda_1'}{\lambda_2'}{k}
  \CG*{k0}{\lambda_1'0}{l0} \CG*{k0}{\lambda_2'0}{L0}
 \,.
\label{eq:G-rho-0}
\end{multline}
Since the charge density operator does not couple to the nucleon spins,
including that amounts to a simple recoupling symbol,
\begin{multline}
 \braket{u;\couple{l}{s}{jm_j}|\tilde{\rho}_{L}(\vecq)%
 |u';\couple{l'}{s'}{j'm_j'}}
 = ({-}1)^{j+s+l'+L} \sqrt{\hat{j}\hat{l'}}
 \CG*{jm_j}{L0}{j'm_j'} \SixJ{l}{s}{j}{j'}{L}{l'} \delta_{ss'}
 \braket{u;l||\tilde{\rho}_{L0}(\vecq)||u';l'} \\
 \equiv \CG*{jm_j}{L0}{j'm_j'}
 \braket{u;\couple{l}{s}{j}||\tilde{\rho}_{L}(q)||u';\couple{l'}{s'}{j'}} \,,
\label{eq:rho-0-jm}
\end{multline}
and overall we have defined, in the second line, the reduced matrix element in
the spin-orbit coupled basis.
\end{widetext}
Finally, the isospin part can be evaluated by writing a two-nucleon coupled
isospin state as
\begin{equation}
 \ket{t m_t} = \sum_{m_1,m_2} \CG*{t_1 m_1}{t_2 m_2}{t m_t}
 \ket{t_1 m_1}\ket{t_2 m_2}
\end{equation}
so that
\begin{multline}
 \Braket{t m_t|\frac{1+\tau_3^{(i)}}{2}|t'm_t'} \\
 = \sum_{m_1,m_2} \CG*{t_1 m_1}{t_2 m_2}{t m_t} \CG*{t_1 m_1}{t_2 m_2}{t' m_t'}
 \frac{1+m_i}{2} \,.
 \label{eqn:matrix_isosp}
\end{multline}

\subsection{Next-to-next-to-leading order charge operator}
\label{sec:CurrentPW-NNLO}

At N2LO, there is a relativistic correction to the LO charge
operator, shown in Eq.~\eqref{eq:rho-2-rel} in the main text.
To derive the partial wave representation of this operator, we begin by
recoupling the spin and spatial parts in spherical tensor form
\begin{multline}
 2i\vecq \cdot \vsigma_1 \times \vecK
 = i\vecq \cdot \vsigma_1 \times (\vecp + \vecp' - \vecq/2) \\
 = \sqrt{6}\left\{
  \sigma_1^1 \otimes [q^1 \otimes (\vec p + \vec p\,' - \vec q/2)^1]^1
 \right\}^{00} \,.
\end{multline}

With Eq.~\eqref{eqn:statebs}, we can decouple the spin and spatial parts as
\begin{widetext}
\begin{multline}
 \braket{u; (ls)j m_j | \left\{
  \sigma_i^1 \otimes [q^1 \otimes (\vec p + \vec p\,' - \vec q/2)^1]^1
 \right\}^{00} | u'; (l's')j' m_j'} \\
 \null = \sum_{\nu \mu} \sum_{m m'} \sum_{m_s m_s'}
 C_{1\nu, 1\mu}^{00} C_{l m, s m_s}^{j m_j}
 C_{l' m', s' m_s'}^{j' m_j'}
 \braket{s m_s| \sigma_i^{1\nu} | s' m_s'}
 \times \braket{
  u; l m |
  [q^1 \otimes (\vec p + \vec p\,' - \vec q/2)^1]^{1\mu} | u'; l m'
 }
\label{eqn:matrix_LS}
\end{multline}
Considering the spatial part, we need to calculate
\begin{multline}
 \braket{u; l m | [q^1 \otimes (\vecp + \vecp' - \vecq/2)^1]^{1\mu}
 \vdelta\big(\vecp-\vecp'-\tfrac12\vecq\big) | u'; l' m'} \\
 \null = \int\dd^3p \int\dd^3p' \, Y^*_{lm}(\hat{\vecp})
 \frac{\delta(u -p)}{u^2} [q^1
 \otimes (\vec p + \vec p\,' - \vec q/2)^1]^{1\mu}
 \delta^{(3)}(\vec p - \vec p\,' - \vec q/2) Y_{l'm'}(\hat{\vecp}')
 \frac{\delta(u'-p')}{u'^2} \\
 \null = \int\dd^3p \int\dd^3p' \, Y^*_{lm}(\hat{\vecp})
 \frac{\delta(u-p)}{u^2}
 \delta^{(3)}(\vec p - \vec p\,' - \vec q/2) Y_{l'm'} (\hat{\vecp}')
 \frac{\delta(u'-p')}{u'^2} \\
 \null \times \frac{4\pi}{3} |\vec q\,|\, | \vec p + \vec p\,' - \vec q/2 |
 \sum_{\lambda_1, \lambda_2=-1}^1 C_{1\lambda_1,1\lambda_2}^{1\mu}
 Y_{1\lambda_1}(\hat{\vecq})
 Y_{1\lambda_2}(\widehat{\vecp + \vecp' - \frac{\vecq}{2}}) \,.
\end{multline}
\end{widetext}

We follow the same procedure as in Sec.~\ref{sec:CurrentPW-LO}, \ie, we
integrate over $\vecp'$ and expand $x \equiv \hat{\vecq} \cdot \hat{\vecp}$
in terms of Legendre polynomials.
We also need the contraction rule for spherical harmonics:
\begin{multline}
 Y_{l'm'}(\widehat{\vecp - \frac{\vecq}{2}})
 Y_{1\lambda_2}(\widehat{\vecp - \frac{\vecq}{2}}) \\
 \null = \sum_{\beta\beta_m}\sqrt{\frac{3\hat{l'}}{4\pi \hat{\beta}}}
 C_{l' 0, 10}^{\beta 0}
 C_{l' m', 1\lambda_2}^{\beta \beta_m}
 Y_{\beta \beta_m}(\widehat{\vecp - \frac{\vecq}{2}}) \,.
\end{multline}
After reducing the product of two coupled spherical harmonics as in
Eq.~\eqref{eqn:conct_biH},
\begin{multline}
 \YY_{\gamma_1 \gamma_2}^{\beta \beta_m}(\hat{\vecp},\hat{\vecq})
 \YY_{kk}^{00}(\hat{\vecp},\hat{\vecq})
 = \frac{1}{4\pi} \sqrt{\hat{k}\hat{\gamma_1}\hat{\gamma_2}}
 (-1)^{\beta+\gamma_1+\gamma_2} \\
 \null \times \sum_{f_1,f_2}
 \SixJ{f_2}{f_1}{\beta}{\gamma_1}{\gamma_2}{k}
 C_{k0,\gamma_1 0}^{f_1 0}
 C_{k0,\gamma_2 0}^{f_2 0}
 \YY^{\beta \beta_m}_{f_1 f_2}(\hat{\vecp},\hat{\vecq}) \,,
\end{multline}
the integral over $ \hat{\vecp}$ can be completed as in Eq.~\eqref{eqn:int_p}:
\begin{multline}
 \int d\Omega_{p} Y_{lm}^*(\hat{\vecp})
 \YY^{\beta \beta_m}_{f_1 f_2}(\hat{\vecp},\hat{\vecq}) \\
 \null = \sum_{m_1 m_2} C_{f_1m_1,f_2m_2}^{\beta \beta_m}
 Y_{f_2m_2}(\hat{\vecq}) \delta_{f_1 l} \delta_{m m_1} \,.
\end{multline}

By collecting these terms together and integrating out $p$,
we arrive at the following:
\begin{widetext}
\begin{multline}
 \braket{u; l m | [q^1 \otimes (\vec p + \vec p\,' - \vec q/2)^1]^{1\mu}
 \vdelta\big(\vecp-\vecp'-\tfrac12\vecq\big)| u'; l' m'} \\
 = \frac{4\pi}{\sqrt{3}} q \sum_{k} \hat{k}(-1)^k
 \int_{-1}^1 \dd x\,P_k(x) \frac{\delta(u' - \iota(u, q, x))}{u'^{2}}
 \sum_{\lambda_1,\lambda_2 = {-}1}^1 C_{1\lambda_1,1\lambda_2}^{1\mu}
 \sum_{\beta\beta_m}\sqrt{\hat{l'}}
 C_{l' 0, 10}^{\beta 0}
 C_{l' m', 1\lambda_2}^{\beta \beta_m}
 \sum_{\gamma_1+\gamma_2 = \beta}
 \frac{u^{\gamma_1}(-\frac{q}{2})^{\gamma_2}}{(\iota(u, q, x))^{\beta-1}} \\
 \null \times \sqrt{\binom{2\beta}{2\gamma_1}}
 \times \sum_{f_2} \SixJ{f_2}{l}{\beta}{\gamma_1}{\gamma_2}{k}
 C_{k0,\gamma_1 0}^{l 0} C_{k0,\gamma_2 0}^{f_2 0}
 \sum_{m_2} C_{l m,f_2m_2}^{\beta \beta_m}
 Y_{f_2m_2}(\hat{\vecq}) Y_{1\lambda_1}(\hat{\vecq}) \,,
\end{multline}
\end{widetext}
where
\begin{equation}
 \iota(p, q, x) = \sqrt{p^2 - p q x + q^2/4} \,.
\label{eq:iota}
\end{equation}
Contracting two spherical harmonics depending on $\hat{\vecq}$, we get
\begin{multline}
 Y_{f_2 m_2}(\hat{\vecq}) Y_{1\lambda_1}(\hat{\vecq}) \\
 \null = \sum_{L m_L}\sqrt{\frac{3\hat{f_2}}{4\pi \hat{L}}}
 C_{f_2 0, 10}^{L 0} C_{f_2 m_2, 1\lambda_1}^{L m_L} Y_{L m_L}(\hat{\vecq}) \,.
\end{multline}
Here $L$ is the rank of multipole, as in Sec.~\ref{sec:CurrentPW-LO}.
Choosing $\vec{q}$ aligned with the $\hat{z}$ axis leads to $m_L = 0$ and
Eq.~\eqref{eqn:YY_qz}.
The sums over $\lambda_1, \lambda_2, \beta_m$ and $m_2$ can be carried out
explicitly by employing relations for sums of products of Clebsch-Gordan
coefficients~\cite{Varshalovich:1988ifq}:
\begin{multline}
 \sum_{\substack{\lambda_1, \lambda_2 \\ \beta_m m_2}}
 C_{1\lambda_1,1\lambda_2}^{1\mu}
 C_{l' m', 1\lambda_2}^{\beta \beta_m}
 C_{l m,f_2m_2}^{\beta \beta_m}
 C_{f_2 m_2, 1\lambda_1}^{L 0} \\
 \null = ({-}1)^{l+l'+f_2 -2\beta}\times ({-}1)^{L - 1 + m -m'} \hspace{5em} \\
 \null \times \hat{\beta}\sqrt{\hat{1}\hat{L}}
 \sum_{k \kappa} C_{l'm',L0}^{k\kappa} C_{lm,1-\mu}^{k\kappa}
 \NineJ%
  {1}{1}{1}%
  {f_2}{\beta}{l}%
  {L}{l'}{k} \,.
\end{multline}

Finally, the spatial part can be simplified as
\begin{widetext}
\begin{multline}
 \braket{u l m | [q^1 \otimes (\vec p + \vec p\,' - \vec q/2)^1]^{1\mu}
 \vdelta\big(\vecp-\vecp'-\tfrac12\vecq\big) | u' l' m'} \\
 \null = q \sqrt{\hat{l'}} \sum_{L} \sqrt{\hat{L}}
 \sum_{k} \hat{k}(-1)^k \int_{-1}^1 dx P_k(x)
 \frac{\delta(u' - \iota(u, q, x))}{u'^{2}}
 \sum_{\beta} C_{l' 0, 10}^{\beta 0}
 \sum_{\gamma_1+\gamma_2 = \beta} \sqrt{\binom{2\beta}{2\gamma_1}}
 \frac{p^{\gamma_1}(-\frac{q}{2})^{\gamma_2}}{(\iota(u, q, x))^{\beta-1}} \\
 \null \times \sum_{f_2} \sqrt{\frac{\hat{f_2}}{\hat{L}}}
 \SixJ{f_2}{l}{\beta}{\gamma_1}{\gamma_2}{k}
 \times  C_{k0,\gamma_1 0}^{l 0} C_{k0,\gamma_2 0}^{f_2 0} C_{f_2 0, 10}^{L 0}
 \times ({-}1)^{l+l'+f_2 -2\beta} \times ({-}1)^{L- 1 + m - m'} \\
 \times \hat{\beta}\sqrt{\hat{1}\hat{L}} \sum_{\bar{k} \kappa}
 C_{l'm',L0}^{\bar{k}\kappa} C_{lm,1 -\mu}^{\bar{k}\kappa}
 \NineJ%
  {1}{1}{1}%
  {f_2}{\beta}{l}%
  {L}{l'}{\bar{k}} \,,
\label{eqn:matrix_orbit_1}
\end{multline}
and the spin part can be evaluated in the same way as the isospin part:
\begin{equation}
 \langle s m_s| \sigma_i^{1\nu} | s' m_s' \rangle
 = ({-}1)^{s'} \sqrt{6\hat{s}'} C_{s'm_s', 1\nu}^{s m_s}
 \SixJ{\frac12}{s}{\frac12}{s'}{\frac12}{1} \,.
\label{eqn:matrix_spin}
\end{equation}

Inserting Eqs.~\eqref{eqn:matrix_orbit_1} and~\eqref{eqn:matrix_spin} into
Eq.~\eqref{eqn:matrix_LS}, and evaluating following sums explicitly:
\begin{multline}
 \sum_{\nu \mu} \sum_{m_l m_l'} \sum_{m_s m_s'} \sum_{\kappa}
 ({-}1)^{m_l - m_l'}
 C_{1\nu, 1\mu}^{00} C_{l m_l, s m_s}^{j m_j}
 C_{l' m_l', s' m_s'}^{j' m_j'} C_{l'm_l',L0}^{\bar{k}\kappa}
 C_{lm_l,1 -\mu}^{\bar{k}\kappa} C_{s'm_s', 1\nu}^{s m_s} \\
 \null = ({-}1)^{l + L + \bar{k} + s' - s + 1} \sqrt{\frac{\hat{j} \hat{s}}{3}}
 C_{jm_j,L0}^{j'm_j'} \,\hat{\bar{k}} \,
 \SixJ{l'}{s'}{j'}{j}{L}{\bar{k}}
 \SixJ{l}{s}{j}{s'}{\bar{k}}{1} \,,
\end{multline}
we arrive at the final result
\begin{multline}
 \braket{u; (ls) j m_j
  | 2i\vec q \cdot \vec \vsigma_i \times \vec K_i \;
    \vdelta\big(\vecp-\vecp'-\tfrac12\vecq\big)
  | u'; (l's')j' m_j'} \\
 \null = 6\sqrt{\hat{s}\hat{s}'} \times ({-}1)^{l'-s} \sqrt{\hat{j} \hat{l'}}
 \SixJ{\frac12}{s}{\frac12}{s'}{\frac12}{1}
 \times q \sum_{L} \sqrt{\hat{L}} C_{jm_j,L0}^{j'm_j'}
 \sum_{k} \hat{k}(-1)^k \int_{-1}^1 dx P_k(x)
 \frac{\delta(u' - \iota(u, q, x))}{u'^{2}} \\
 \null \times \sum_{\beta} \hat{\beta} \sum_{\gamma_1+\gamma_2 = \beta}
 \sqrt{\binom{2\beta}{2\gamma_1}}
 \frac{p^{\gamma_1}(-\frac{q}{2})^{\gamma_2}}{(\iota(u, q, x))^{\beta-1}}
 \sum_{f_2} ({-}1)^{f_2} \sqrt{\hat{f_2}}
 \SixJ{f_2}{l}{\beta}{\gamma_1}{\gamma_2}{k}
 C_{l' 0, 10}^{\beta 0} C_{k0,\gamma_1 0}^{l 0}
 C_{k0,\gamma_2 0}^{f_2 0} C_{f_2 0, 10}^{L 0} \\
 \null \times \sum_{\bar{k}} ({-}1)^{\bar{k}} \, \hat{\bar{k}} \,
 \SixJ{l'}{s'}{j'}{j}{L}{\bar{k}}
 \SixJ{l}{s}{j}{s'}{\bar{k}}{1}
 \NineJ%
  {1}{1}{1}%
  {f_2}{\beta}{l}%
  {L}{l'}{\bar{k}} \,.
\label{eq:rho-2-jm}
\end{multline}
The isospin part can be evaluated easily as in Eq.~\eqref{eqn:matrix_isosp}.
\end{widetext}

\subsection{Effective boost-correction charge operator}
\label{sec:CurrentPW-Boost}

As mentioned in Sec.~\ref{sec:CurrentOp-Boost}, the corrections that arise from
boosting the initial deuteron state from the lab frame into the final-state
center-of-mass frame, which enter at N2LO, can be written in terms of effective
corrections to the current operator.
To derive these, we start from Eq.~\eqref{eq:rho-0-G} and consider the
subsitution $q \to {q}/{\sqrt{1+\eta}}$.
Expanding the result up to first order in $\eta=q^2/(4 M_d^2)$ leads to the
following matrix element:
\begin{widetext}
\begin{multline}
 \braket{u;lm|\rho^{(2)}_{\text{boost}}(\vecq)|u';l'm'}
 = \sum_{L} \sqrt{\hat{L}} \, C_{l m,L 0}^{l'm'} \int_{-1}^1 \dd x
 \, \eta \times \Bigg[
  G^{L,\text{boost,a}}_{l,l'}(u,q,x)
  \frac{\delta'\big(u'-\iota(u,q,x)\big)}{u'^2} \\
  \null + G^{L,\text{boost,b}}_{l,l'}(u,q,x)
  \frac{\delta\big(u'-\iota(u,q,x)\big)}{u'^2}
 \Bigg] \,,
\label{eq:rho-0-G-boost}
\end{multline}
where
\begin{subalign}
 G^{L,\text{boost,a}}_{l,l'}(u,q,x)
 &= \frac{G^{L}_{l,l'}(u,q,x)}{\iota(u,q,x)^2}\left(
 \frac{q^2}{8}-\frac{u q x}{4}\right) \,, \\
 G^{L,\text{boost,b}}_{l,l'}(u,q,x)
 &= \frac{l' G^{L}_{l,l'}(u,q,x)}{\iota(u,q,x)}\left(
 \frac{q^2}{8}-\frac{u q x}{4}\right)
 - \frac{\tilde{G}^{L\text{boost,b}}_{l,l'}(u,q,x)}{\iota(u,q,x)} \,,
\end{subalign}
and
\begin{multline}
 \tilde{G}^{L\text{boost,b}}_{l,l'}(u,q,x)
 = \frac12 \sum_{k} \sqrt{\hat{k}} ({-}1)^k P_k(x)
 \sum_{\lambda_1'+\lambda_2'=l'}
  \frac{u^{\lambda_1'}\big({-}\tfrac12q\big)^{\lambda_2'}}
  {\iota(u,q,x)^{l'}}
  \frac{\lambda_2'}{2}
  \sqrt{\frac{(2l'+1)!}{(2\lambda_1'+1)!(2\lambda_2'+1)!}}
 \sqrt{\hat{k}\hat{\lambda_1'}\hat{\lambda_2'}}
  ({-}1)^{\lambda_1'+\lambda_2'+l'} \\
 \null \times \SixJ{L}{l}{l'}{\lambda_1'}{\lambda_2'}{k}
  \CG*{k0}{\lambda_1'0}{l0} \CG*{k0}{\lambda_2'0}{L0}
 \,.
\end{multline}
\end{widetext}

\section{Inversion of the LIT}
\label{sec:Inversion}

To invert the LIT, we follow the basic procedure discussed in
Ref.~\cite{Efros:2007nq}, with certain extensions that we describe in this
section.
The basic idea of the approach is to expand the response function over a set of
functions $\chi_{n}(\omega)$,
\begin{equation}
\label{eq:fitResp}
 \mathcal{R}(\omega) = \sum_{n}^{N} \alpha_{n} \chi_{n}(\omega) \,,
\end{equation}
where $\alpha_{n}$ are coefficients to be determined.
The precise form chosen for the $\chi_{n}(\omega)$ will be given below.
Since the LIT as a mapping between functions is a linear operation, combining
Eq.~\eqref{eq:fitResp} with Eq.~\eqref{eq:LIT} gives a related expansion for the
LIT of the response function,
\begin{equation}
\label{eq:fitLIT}
 \Phi(\sigma_{R}) = \sum_{n}^{N} \alpha_{n} \Tilde{\chi}_{n}(\sigma_{R}) \,,
\end{equation}
where the $\Tilde{\chi_{n}}$ are given by
\begin{equation}
 \Tilde{\chi}_{n}(\sigma_{R})
 = \int_{\omega_{\text{th}}}^{\infty} d\omega
 \frac{\chi_{n}(\omega)}{(\omega - \sigma_{R})^2+\sigma_{I}^{2}} \,.
\end{equation}
That is, $\Tilde{\chi}_{n}(\sigma_{R})$ is just the LIT of $\chi_{n}(\omega)$,
and the inversion is carried out by fitting Eq.~\eqref{eq:fitLIT} to the LIT
obtained as described in the main text.
This procedure, while straightforward in principle, is still delicate in
practice to the ill-posedness of the LIT inversion problem (see
Sec.~\ref{sec:ExpData}).

For the $\chi_{n}(\omega)$ it is important to choose a form that (a) is able to
capture the actual structure of the response function and (b) ensures that the
fit does not suffer from strong correlations among different $\chi_{n}(\omega)$.
We further follow Ref.~\cite{Efros:2007nq} in this regard and set
\begin{equation}
 \label{eq:chi-n}
 \chi_{n}(\omega) = \epsilon^{\beta_0} \ee^{{-}\epsilon \beta_{1}/n} \,,
\end{equation}
with $\epsilon = \omega - \omega_{\text{th}}$, and with $\beta_{0}$ and
$\beta_{1}$ to be treated as nonlinear fit parameters.
While Ref.~\cite{Efros:2007nq} suggests to fix $\beta_0$ based on physics
considerations (enforcing a particular energy dependence at threshold), we find
it more useful to let both $\beta_0$ and $\beta_1$ vary along with the
linear fit parameters $\alpha_n$.
Moreover, also the optimal number $N$ of terms in Eqs.~\eqref{eq:fitResp}
and~\eqref{eq:fitLIT} has to be determined empirically as part of the fitting
procedure -- too few terms will in general give a poor fit, but too many terms
may lead to high-frequency oscillations and therefore unstable fits.

Instead of treating $\beta_0$ and $\beta_1$ on equal footing with the $\alpha_n$
in a combined non-linear fit, we chose to sample the two non-linear parameters
across a relatively wide range, and then we run a linear fit for each sample of
$(\beta_0,\beta_1)$ to determine the corresponding $\alpha_n$.
For the linear fits, we perform Ridge regression (a regularized form of least
squares) and ultimately calculate the linear coefficients using singular value
decomposition (see, \eg in Ref.~\cite{HjorthJensen:2023}, Chapter 4).
This leads to an overall large collection of possible inversion results, many of
which can be ruled out for poor performance (indicated by large fit residuals).
Among the inversion candidates not ruled out in this way, there can still be
some that feature implausible high-frequency oscillations.
As discussed, this is, in general, a manifestation of the ill-posedness of the
LIT inversion.
We therefore choose to further filter the results by their behavior in Fourier
space -- implemented as a fast Fourier transform (FFT) -- using the presence of
significant power in the spectrum above a threshold frequency defined by the
inverse width of the Lorentz kernel, $\xi_{\text{th}}=1/2\pi\sigma_I$ , as a
natural criterion to discount a candidate inversion.
This particular weight we use is the integral of the Fourier transform of the
candidate response from $\xi_{\text{th}}$ to infinity.
As we are using the FFT, this is implemented in practice as a discrete sum of
Fourier components between $\xi_{\text{th}}$ and the maximum frequency
determined by the discrete mesh used for $\omega$ (which, in turn, is determined
by the $\sigma_R$ mesh that the LIT was calculated on).

In practice we do not use a strict acceptance/rejection but rather sort all
candidates by their performance -- with respect to both fit residuals and
high-frequency power.
Specifically, we found the following prescription to work
well:
\begin{enumerate}
 \item Out of the initial set of candidates from sampling and fitting, we keep
  the best $N_{\text{initial}}$ out of $N_{total}$ fits.
 \item Those candidates we then sort by their Fourier weight, as defined above,
  and keep the best $N_\text{Fourier}$ responses.
 \item Out of these, we pick again the best $N_{\text{final}}$ to calculate our
  final result for the response as the mean among those remaining candidates,
  with an uncertainty band provided by calculating the 97.5\% percentile around
  the mean.
\end{enumerate}
Reasonable values for $N_{\text{initial}}$, $N_{\text{Fourier}}$, and
$N_{\text{final}}$, in addition to the number $N$ of terms in
Eq.~\eqref{eq:fitResp}, still need to be determined empirically, but overall the
inversion proceeds in an automatic fashion, and in practice one needs only
ensure that all these numbers are large enough for the final result to become
effectively independent of the particular choice.

\begin{figure}[tbhp]
 \centering
 \includegraphics[scale=0.5]{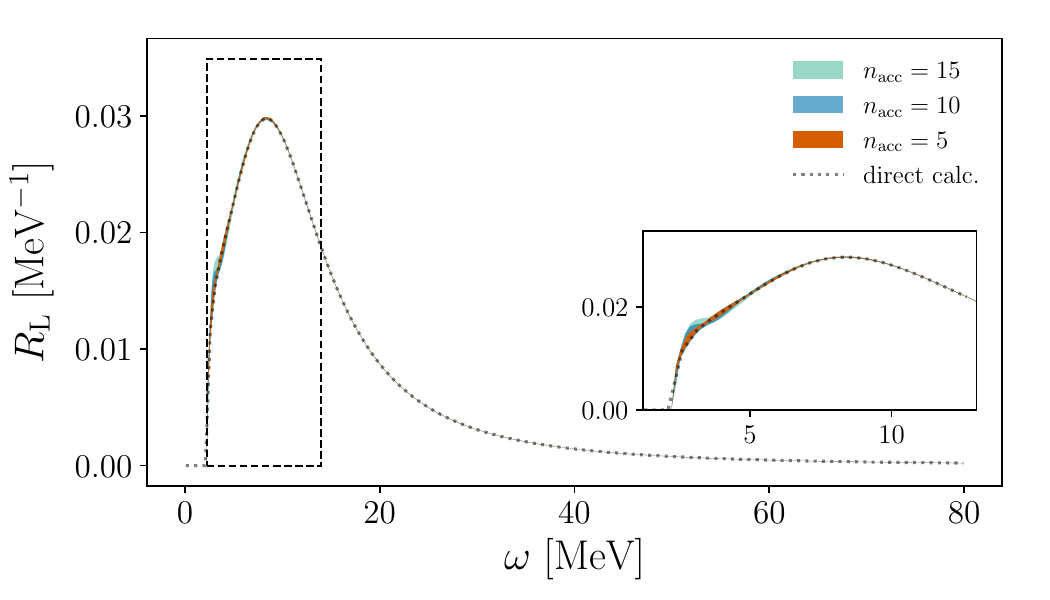}
 \caption{
  Deuteron response from the inverted LIT for $\Lambda = 800$ MeV,
  $\sigma_I = 3.0~\MeV$ for momentum transfer
  $\vecq^2 = 0.6$ fm$^{-2}$.
  \label{fig:exact-0p6}
 }
\end{figure}

\begin{figure}[tbhp]
 \centering
 \includegraphics[scale=0.5]{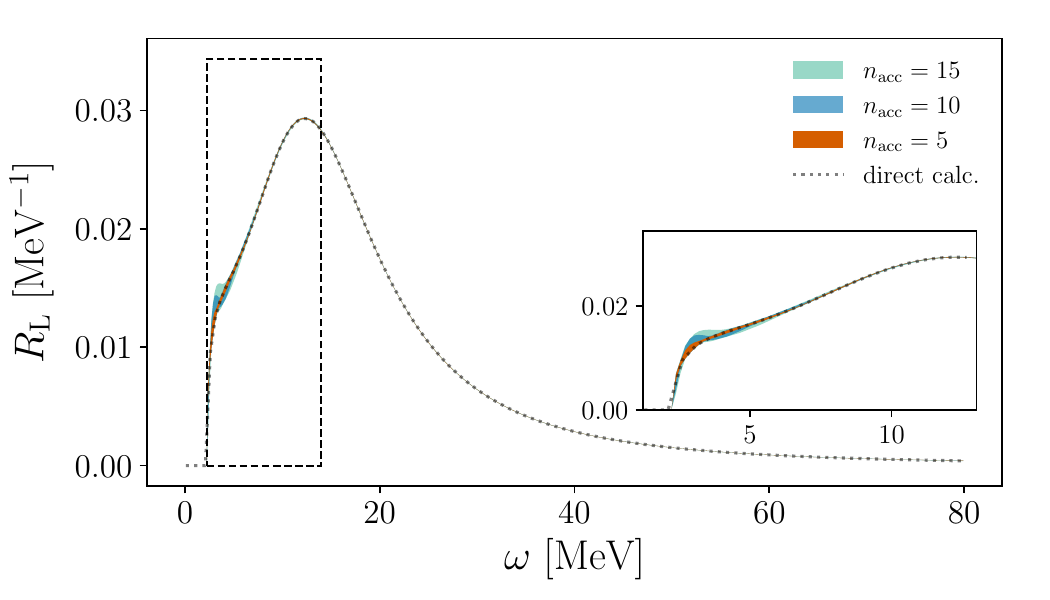}
 \caption{
  Deuteron response from the inverted LIT $\Lambda = 800$ MeV,
  $\sigma_I = 3.0~\MeV$ for momentum transfer
  $\vecq^2 = 1.0$ fm$^{-2}$.
  \label{fig:exact-1p0}
 }
\end{figure}

As mentioned in the main text, a direct calculation of the deuteron response
function is in fact a tractable task, which we have implemented at leading order
in the EFT expansion.
We can therefore use exact LO results to benchmark our inversion method.
In Figs.~\ref{fig:exact-0p6},~\ref{fig:exact-1p0} we show results obtained from
inverting LO LIT calculations at $\sigma_I = 3.0,5.0,10.0~\MeV$ compared to a
direct calculation of the LO response function.

The values used to obtain the results presented in Fig.\ref{fig:exact-0p6},
\ref{fig:exact-1p0} are $N=15$, $N_{\text{initial}}=200$, $N_{\text{Fourier}}=20$ and
$N_{\text{Final}} = 15, 10, 5$ as indicated in the legend.
If one chooses smaller values for $N_{\text{initial}}$, there is a tendency to
eliminate many of the smoother functions and one is left with only highly
oscillatory results that are deemed unphysical for reasons discussed above.
If one sorts by Fourier weight first, then the top-ranked results have relatively
large residuals.
The procedure remains independent of knowledge of the exact response function,
which is important for cases where direct verification of inversion is unfeasible.


\bibliographystyle{apsrev4-1}

\begin{thebibliography}{64}%
\makeatletter
\providecommand \@ifxundefined [1]{%
 \@ifx{#1\undefined}
}%
\providecommand \@ifnum [1]{%
 \ifnum #1\expandafter \@firstoftwo
 \else \expandafter \@secondoftwo
 \fi
}%
\providecommand \@ifx [1]{%
 \ifx #1\expandafter \@firstoftwo
 \else \expandafter \@secondoftwo
 \fi
}%
\providecommand \natexlab [1]{#1}%
\providecommand \enquote  [1]{``#1''}%
\providecommand \bibnamefont  [1]{#1}%
\providecommand \bibfnamefont [1]{#1}%
\providecommand \citenamefont [1]{#1}%
\providecommand \href@noop [0]{\@secondoftwo}%
\providecommand \href [0]{\begingroup \@sanitize@url \@href}%
\providecommand \@href[1]{\@@startlink{#1}\@@href}%
\providecommand \@@href[1]{\endgroup#1\@@endlink}%
\providecommand \@sanitize@url [0]{\catcode `\\12\catcode `\$12\catcode
  `\&12\catcode `\#12\catcode `\^12\catcode `\_12\catcode `\%12\relax}%
\providecommand \@@startlink[1]{}%
\providecommand \@@endlink[0]{}%
\providecommand \url  [0]{\begingroup\@sanitize@url \@url }%
\providecommand \@url [1]{\endgroup\@href {#1}{\urlprefix }}%
\providecommand \urlprefix  [0]{URL }%
\providecommand \Eprint [0]{\href }%
\providecommand \doibase [0]{http://dx.doi.org/}%
\providecommand \selectlanguage [0]{\@gobble}%
\providecommand \bibinfo  [0]{\@secondoftwo}%
\providecommand \bibfield  [0]{\@secondoftwo}%
\providecommand \translation [1]{[#1]}%
\providecommand \BibitemOpen [0]{}%
\providecommand \bibitemStop [0]{}%
\providecommand \bibitemNoStop [0]{.\EOS\space}%
\providecommand \EOS [0]{\spacefactor3000\relax}%
\providecommand \BibitemShut  [1]{\csname bibitem#1\endcsname}%
\let\auto@bib@innerbib\@empty
\bibitem [{\citenamefont {Hammer}\ \emph {et~al.}(2020)\citenamefont {Hammer},
  \citenamefont {K{\"o}nig},\ and\ \citenamefont {{van
  Kolck}}}]{Hammer:2019poc}%
  \BibitemOpen
  \bibfield  {author} {\bibinfo {author} {\bibfnamefont {H.-W.}\ \bibnamefont
  {Hammer}}, \bibinfo {author} {\bibfnamefont {S.}~\bibnamefont {K{\"o}nig}}, \
  and\ \bibinfo {author} {\bibfnamefont {U.}~\bibnamefont {{van Kolck}}},\
  }\href {\doibase 10.1103/RevModPhys.92.025004} {\bibfield  {journal}
  {\bibinfo  {journal} {Rev. Mod. Phys.}\ }\textbf {\bibinfo {volume} {92}},\
  \bibinfo {pages} {025004} (\bibinfo {year} {2020})}\BibitemShut {NoStop}%
\bibitem [{\citenamefont {Weinberg}(1990)}]{Weinberg:1990rz}%
  \BibitemOpen
  \bibfield  {author} {\bibinfo {author} {\bibfnamefont {S.}~\bibnamefont
  {Weinberg}},\ }\href {\doibase 10.1016/0370-2693(90)90938-3} {\bibfield
  {journal} {\bibinfo  {journal} {Phys. Lett. B}\ }\textbf {\bibinfo {volume}
  {251}},\ \bibinfo {pages} {288} (\bibinfo {year} {1990})}\BibitemShut
  {NoStop}%
\bibitem [{\citenamefont {Weinberg}(1991)}]{Weinberg:1991um}%
  \BibitemOpen
  \bibfield  {author} {\bibinfo {author} {\bibfnamefont {S.}~\bibnamefont
  {Weinberg}},\ }\href {\doibase 10.1016/0550-3213(91)90231-L} {\bibfield
  {journal} {\bibinfo  {journal} {Nucl. Phys. B}\ }\textbf {\bibinfo {volume}
  {363}},\ \bibinfo {pages} {3} (\bibinfo {year} {1991})}\BibitemShut {NoStop}%
\bibitem [{\citenamefont {Ordonez}\ and\ \citenamefont {van
  Kolck}(1992)}]{Ordonez:1992xp}%
  \BibitemOpen
  \bibfield  {author} {\bibinfo {author} {\bibfnamefont {C.}~\bibnamefont
  {Ordonez}}\ and\ \bibinfo {author} {\bibfnamefont {U.}~\bibnamefont {van
  Kolck}},\ }\href {\doibase 10.1016/0370-2693(92)91404-W} {\bibfield
  {journal} {\bibinfo  {journal} {Phys. Lett. B}\ }\textbf {\bibinfo {volume}
  {291}},\ \bibinfo {pages} {459} (\bibinfo {year} {1992})}\BibitemShut
  {NoStop}%
\bibitem [{\citenamefont {Weinberg}(1992)}]{Weinberg:1992yk}%
  \BibitemOpen
  \bibfield  {author} {\bibinfo {author} {\bibfnamefont {S.}~\bibnamefont
  {Weinberg}},\ }\href {\doibase 10.1016/0370-2693(92)90099-P} {\bibfield
  {journal} {\bibinfo  {journal} {Phys. Lett. B}\ }\textbf {\bibinfo {volume}
  {295}},\ \bibinfo {pages} {114} (\bibinfo {year} {1992})},\ \Eprint
  {http://arxiv.org/abs/hep-ph/9209257} {arXiv:hep-ph/9209257} \BibitemShut
  {NoStop}%
\bibitem [{\citenamefont {Van~Kolck}(1993)}]{vanKolck:1993}%
  \BibitemOpen
  \bibfield  {author} {\bibinfo {author} {\bibfnamefont {U.~L.}\ \bibnamefont
  {Van~Kolck}},\ }\emph {\bibinfo {title} {{Soft Physics: Applications of
  Effective Chiral Lagrangians to Nuclear Physics and Quark Models}}},\
  \href@noop {} {Ph.D. thesis},\ \bibinfo  {school} {Texas U.} (\bibinfo {year}
  {1993})\BibitemShut {NoStop}%
\bibitem [{\citenamefont {Rho}(1991)}]{Rho:1990cf}%
  \BibitemOpen
  \bibfield  {author} {\bibinfo {author} {\bibfnamefont {M.}~\bibnamefont
  {Rho}},\ }\href {\doibase 10.1103/PhysRevLett.66.1275} {\bibfield  {journal}
  {\bibinfo  {journal} {Phys. Rev. Lett.}\ }\textbf {\bibinfo {volume} {66}},\
  \bibinfo {pages} {1275} (\bibinfo {year} {1991})}\BibitemShut {NoStop}%
\bibitem [{\citenamefont {Park}\ \emph {et~al.}(1993)\citenamefont {Park},
  \citenamefont {Min},\ and\ \citenamefont {Rho}}]{Park:1993jf}%
  \BibitemOpen
  \bibfield  {author} {\bibinfo {author} {\bibfnamefont {T.-S.}\ \bibnamefont
  {Park}}, \bibinfo {author} {\bibfnamefont {D.-P.}\ \bibnamefont {Min}}, \
  and\ \bibinfo {author} {\bibfnamefont {M.}~\bibnamefont {Rho}},\ }\href
  {\doibase 10.1016/0370-1573(93)90099-Y} {\bibfield  {journal} {\bibinfo
  {journal} {Phys. Rept.}\ }\textbf {\bibinfo {volume} {233}},\ \bibinfo
  {pages} {341} (\bibinfo {year} {1993})},\ \Eprint
  {http://arxiv.org/abs/hep-ph/9301295} {arXiv:hep-ph/9301295} \BibitemShut
  {NoStop}%
\bibitem [{\citenamefont {Park}\ \emph {et~al.}(1996)\citenamefont {Park},
  \citenamefont {Min},\ and\ \citenamefont {Rho}}]{Park:1995pn}%
  \BibitemOpen
  \bibfield  {author} {\bibinfo {author} {\bibfnamefont {T.-S.}\ \bibnamefont
  {Park}}, \bibinfo {author} {\bibfnamefont {D.-P.}\ \bibnamefont {Min}}, \
  and\ \bibinfo {author} {\bibfnamefont {M.}~\bibnamefont {Rho}},\ }\href
  {\doibase 10.1016/0375-9474(95)00406-8} {\bibfield  {journal} {\bibinfo
  {journal} {Nucl. Phys. A}\ }\textbf {\bibinfo {volume} {596}},\ \bibinfo
  {pages} {515} (\bibinfo {year} {1996})},\ \Eprint
  {http://arxiv.org/abs/nucl-th/9505017} {arXiv:nucl-th/9505017} \BibitemShut
  {NoStop}%
\bibitem [{\citenamefont {K{\"o}lling}\ \emph {et~al.}(2009)\citenamefont
  {K{\"o}lling}, \citenamefont {Epelbaum}, \citenamefont {Krebs},\ and\
  \citenamefont {Mei{\ss}ner}}]{Kolling:2009iq}%
  \BibitemOpen
  \bibfield  {author} {\bibinfo {author} {\bibfnamefont {S.}~\bibnamefont
  {K{\"o}lling}}, \bibinfo {author} {\bibfnamefont {E.}~\bibnamefont
  {Epelbaum}}, \bibinfo {author} {\bibfnamefont {H.}~\bibnamefont {Krebs}}, \
  and\ \bibinfo {author} {\bibfnamefont {U.-G.}\ \bibnamefont {Mei{\ss}ner}},\
  }\href {\doibase 10.1103/PhysRevC.80.045502} {\bibfield  {journal} {\bibinfo
  {journal} {Phys. Rev. C}\ }\textbf {\bibinfo {volume} {80}},\ \bibinfo
  {pages} {045502} (\bibinfo {year} {2009})},\ \Eprint
  {http://arxiv.org/abs/0907.3437} {arXiv:0907.3437 [nucl-th]} \BibitemShut
  {NoStop}%
\bibitem [{\citenamefont {K{\"o}lling}\ \emph {et~al.}(2011)\citenamefont
  {K{\"o}lling}, \citenamefont {Epelbaum}, \citenamefont {Krebs},\ and\
  \citenamefont {Mei{\ss}ner}}]{Kolling:2011mt}%
  \BibitemOpen
  \bibfield  {author} {\bibinfo {author} {\bibfnamefont {S.}~\bibnamefont
  {K{\"o}lling}}, \bibinfo {author} {\bibfnamefont {E.}~\bibnamefont
  {Epelbaum}}, \bibinfo {author} {\bibfnamefont {H.}~\bibnamefont {Krebs}}, \
  and\ \bibinfo {author} {\bibfnamefont {U.-G.}\ \bibnamefont {Mei{\ss}ner}},\
  }\href {\doibase 10.1103/PhysRevC.84.054008} {\bibfield  {journal} {\bibinfo
  {journal} {Phys. Rev. C}\ }\textbf {\bibinfo {volume} {84}},\ \bibinfo
  {pages} {054008} (\bibinfo {year} {2011})},\ \Eprint
  {http://arxiv.org/abs/1107.0602} {arXiv:1107.0602 [nucl-th]} \BibitemShut
  {NoStop}%
\bibitem [{\citenamefont {Pastore}\ \emph {et~al.}(2011)\citenamefont
  {Pastore}, \citenamefont {Girlanda}, \citenamefont {Schiavilla},\ and\
  \citenamefont {Viviani}}]{Pastore:2011ip}%
  \BibitemOpen
  \bibfield  {author} {\bibinfo {author} {\bibfnamefont {S.}~\bibnamefont
  {Pastore}}, \bibinfo {author} {\bibfnamefont {L.}~\bibnamefont {Girlanda}},
  \bibinfo {author} {\bibfnamefont {R.}~\bibnamefont {Schiavilla}}, \ and\
  \bibinfo {author} {\bibfnamefont {M.}~\bibnamefont {Viviani}},\ }\href
  {\doibase 10.1103/PhysRevC.84.024001} {\bibfield  {journal} {\bibinfo
  {journal} {Phys. Rev. C}\ }\textbf {\bibinfo {volume} {84}},\ \bibinfo
  {pages} {024001} (\bibinfo {year} {2011})},\ \Eprint
  {http://arxiv.org/abs/1106.4539} {arXiv:1106.4539 [nucl-th]} \BibitemShut
  {NoStop}%
\bibitem [{\citenamefont {Baroni}\ \emph {et~al.}(2016)\citenamefont {Baroni},
  \citenamefont {Girlanda}, \citenamefont {Pastore}, \citenamefont
  {Schiavilla},\ and\ \citenamefont {Viviani}}]{Baroni:2015uza}%
  \BibitemOpen
  \bibfield  {author} {\bibinfo {author} {\bibfnamefont {A.}~\bibnamefont
  {Baroni}}, \bibinfo {author} {\bibfnamefont {L.}~\bibnamefont {Girlanda}},
  \bibinfo {author} {\bibfnamefont {S.}~\bibnamefont {Pastore}}, \bibinfo
  {author} {\bibfnamefont {R.}~\bibnamefont {Schiavilla}}, \ and\ \bibinfo
  {author} {\bibfnamefont {M.}~\bibnamefont {Viviani}},\ }\href {\doibase
  10.1103/PhysRevC.93.049902} {\bibfield  {journal} {\bibinfo  {journal} {Phys.
  Rev. C}\ }\textbf {\bibinfo {volume} {93}},\ \bibinfo {pages} {015501}
  (\bibinfo {year} {2016})},\ \bibinfo {note} {[Erratum: Phys.Rev.C 93, 049902
  (2016), Erratum: Phys.Rev.C 95, 059901 (2017)]},\ \Eprint
  {http://arxiv.org/abs/1509.07039} {arXiv:1509.07039 [nucl-th]} \BibitemShut
  {NoStop}%
\bibitem [{\citenamefont {Krebs}(2020)}]{Krebs:2020pii}%
  \BibitemOpen
  \bibfield  {author} {\bibinfo {author} {\bibfnamefont {H.}~\bibnamefont
  {Krebs}},\ }\href {\doibase 10.1140/epja/s10050-020-00230-9} {\bibfield
  {journal} {\bibinfo  {journal} {Eur. Phys. J. A}\ }\textbf {\bibinfo {volume}
  {56}},\ \bibinfo {pages} {234} (\bibinfo {year} {2020})},\ \Eprint
  {http://arxiv.org/abs/2008.00974} {arXiv:2008.00974 [nucl-th]} \BibitemShut
  {NoStop}%
\bibitem [{\citenamefont {Hoferichter}\ \emph {et~al.}(2015)\citenamefont
  {Hoferichter}, \citenamefont {Klos},\ and\ \citenamefont
  {Schwenk}}]{Hoferichter:2015ipa}%
  \BibitemOpen
  \bibfield  {author} {\bibinfo {author} {\bibfnamefont {M.}~\bibnamefont
  {Hoferichter}}, \bibinfo {author} {\bibfnamefont {P.}~\bibnamefont {Klos}}, \
  and\ \bibinfo {author} {\bibfnamefont {A.}~\bibnamefont {Schwenk}},\ }\href
  {\doibase 10.1016/j.physletb.2015.05.041} {\bibfield  {journal} {\bibinfo
  {journal} {Phys. Lett. B}\ }\textbf {\bibinfo {volume} {746}},\ \bibinfo
  {pages} {410} (\bibinfo {year} {2015})}\BibitemShut {NoStop}%
\bibitem [{\citenamefont {Cirigliano}\ \emph {et~al.}(2018)\citenamefont
  {Cirigliano}, \citenamefont {Dekens}, \citenamefont {{de Vries}},
  \citenamefont {Graesser},\ and\ \citenamefont
  {Mereghetti}}]{Cirigliano:2018yza}%
  \BibitemOpen
  \bibfield  {author} {\bibinfo {author} {\bibfnamefont {V.}~\bibnamefont
  {Cirigliano}}, \bibinfo {author} {\bibfnamefont {W.}~\bibnamefont {Dekens}},
  \bibinfo {author} {\bibfnamefont {J.}~\bibnamefont {{de Vries}}}, \bibinfo
  {author} {\bibfnamefont {M.~L.}\ \bibnamefont {Graesser}}, \ and\ \bibinfo
  {author} {\bibfnamefont {E.}~\bibnamefont {Mereghetti}},\ }\href {\doibase
  10.1007/JHEP12(2018)097} {\bibfield  {journal} {\bibinfo  {journal} {JHEP}\
  }\textbf {\bibinfo {volume} {2018}},\ \bibinfo {pages} {97} (\bibinfo {year}
  {2018})}\BibitemShut {NoStop}%
\bibitem [{\citenamefont {Oosterhof}\ \emph {et~al.}(2019)\citenamefont
  {Oosterhof}, \citenamefont {Long}, \citenamefont {de~Vries}, \citenamefont
  {Timmermans},\ and\ \citenamefont {van Kolck}}]{Oosterhof:2019dlo}%
  \BibitemOpen
  \bibfield  {author} {\bibinfo {author} {\bibfnamefont {F.}~\bibnamefont
  {Oosterhof}}, \bibinfo {author} {\bibfnamefont {B.}~\bibnamefont {Long}},
  \bibinfo {author} {\bibfnamefont {J.}~\bibnamefont {de~Vries}}, \bibinfo
  {author} {\bibfnamefont {R.~G.~E.}\ \bibnamefont {Timmermans}}, \ and\
  \bibinfo {author} {\bibfnamefont {U.}~\bibnamefont {van Kolck}},\ }\href
  {\doibase 10.1103/PhysRevLett.122.172501} {\bibfield  {journal} {\bibinfo
  {journal} {Phys. Rev. Lett.}\ }\textbf {\bibinfo {volume} {122}},\ \bibinfo
  {pages} {172501} (\bibinfo {year} {2019})},\ \Eprint
  {http://arxiv.org/abs/1902.05342} {arXiv:1902.05342 [hep-ph]} \BibitemShut
  {NoStop}%
\bibitem [{\citenamefont {Liu}\ \emph {et~al.}(2022)\citenamefont {Liu},
  \citenamefont {Peng}, \citenamefont {Lyu},\ and\ \citenamefont
  {Long}}]{Liu:2022cfd}%
  \BibitemOpen
  \bibfield  {author} {\bibinfo {author} {\bibfnamefont {T.-X.}\ \bibnamefont
  {Liu}}, \bibinfo {author} {\bibfnamefont {R.}~\bibnamefont {Peng}}, \bibinfo
  {author} {\bibfnamefont {S.}~\bibnamefont {Lyu}}, \ and\ \bibinfo {author}
  {\bibfnamefont {B.}~\bibnamefont {Long}},\ }\href {\doibase
  10.1103/PhysRevC.106.055501} {\bibfield  {journal} {\bibinfo  {journal}
  {Phys. Rev. C}\ }\textbf {\bibinfo {volume} {106}},\ \bibinfo {pages}
  {055501} (\bibinfo {year} {2022})},\ \Eprint
  {http://arxiv.org/abs/2207.04241} {arXiv:2207.04241 [nucl-th]} \BibitemShut
  {NoStop}%
\bibitem [{\citenamefont {Nogga}\ \emph {et~al.}(2005)\citenamefont {Nogga},
  \citenamefont {Timmermans},\ and\ \citenamefont {van Kolck}}]{Nogga:2005hy}%
  \BibitemOpen
  \bibfield  {author} {\bibinfo {author} {\bibfnamefont {A.}~\bibnamefont
  {Nogga}}, \bibinfo {author} {\bibfnamefont {R.~G.~E.}\ \bibnamefont
  {Timmermans}}, \ and\ \bibinfo {author} {\bibfnamefont {U.}~\bibnamefont {van
  Kolck}},\ }\href {\doibase 10.1103/PhysRevC.72.054006} {\bibfield  {journal}
  {\bibinfo  {journal} {Phys. Rev. C}\ }\textbf {\bibinfo {volume} {72}},\
  \bibinfo {pages} {054006} (\bibinfo {year} {2005})},\ \Eprint
  {http://arxiv.org/abs/nucl-th/0506005} {arXiv:nucl-th/0506005} \BibitemShut
  {NoStop}%
\bibitem [{\citenamefont {Pavon~Valderrama}(2011)}]{PavonValderrama:2011fcz}%
  \BibitemOpen
  \bibfield  {author} {\bibinfo {author} {\bibfnamefont {M.}~\bibnamefont
  {Pavon~Valderrama}},\ }\href {\doibase 10.1103/PhysRevC.84.064002} {\bibfield
   {journal} {\bibinfo  {journal} {Phys. Rev. C}\ }\textbf {\bibinfo {volume}
  {84}},\ \bibinfo {pages} {064002} (\bibinfo {year} {2011})},\ \Eprint
  {http://arxiv.org/abs/1108.0872} {arXiv:1108.0872 [nucl-th]} \BibitemShut
  {NoStop}%
\bibitem [{\citenamefont {Long}\ and\ \citenamefont
  {Yang}(2011)}]{Long:2011qx}%
  \BibitemOpen
  \bibfield  {author} {\bibinfo {author} {\bibfnamefont {B.}~\bibnamefont
  {Long}}\ and\ \bibinfo {author} {\bibfnamefont {C.~J.}\ \bibnamefont
  {Yang}},\ }\href {\doibase 10.1103/PhysRevC.84.057001} {\bibfield  {journal}
  {\bibinfo  {journal} {Phys. Rev. C}\ }\textbf {\bibinfo {volume} {84}},\
  \bibinfo {pages} {057001} (\bibinfo {year} {2011})},\ \Eprint
  {http://arxiv.org/abs/1108.0985} {arXiv:1108.0985 [nucl-th]} \BibitemShut
  {NoStop}%
\bibitem [{\citenamefont {Shi}\ \emph {et~al.}(2022)\citenamefont {Shi},
  \citenamefont {Peng}, \citenamefont {Liu}, \citenamefont {Lyu},\ and\
  \citenamefont {Long}}]{Shi:2022blm}%
  \BibitemOpen
  \bibfield  {author} {\bibinfo {author} {\bibfnamefont {W.}~\bibnamefont
  {Shi}}, \bibinfo {author} {\bibfnamefont {R.}~\bibnamefont {Peng}}, \bibinfo
  {author} {\bibfnamefont {T.-X.}\ \bibnamefont {Liu}}, \bibinfo {author}
  {\bibfnamefont {S.}~\bibnamefont {Lyu}}, \ and\ \bibinfo {author}
  {\bibfnamefont {B.}~\bibnamefont {Long}},\ }\href {\doibase
  10.1103/PhysRevC.106.015505} {\bibfield  {journal} {\bibinfo  {journal}
  {Phys. Rev. C}\ }\textbf {\bibinfo {volume} {106}},\ \bibinfo {pages}
  {015505} (\bibinfo {year} {2022})}\BibitemShut {NoStop}%
\bibitem [{\citenamefont {Long}\ and\ \citenamefont
  {Yang}(2012{\natexlab{a}})}]{Long:2011xw}%
  \BibitemOpen
  \bibfield  {author} {\bibinfo {author} {\bibfnamefont {B.}~\bibnamefont
  {Long}}\ and\ \bibinfo {author} {\bibfnamefont {C.~J.}\ \bibnamefont
  {Yang}},\ }\href {\doibase 10.1103/PhysRevC.85.034002} {\bibfield  {journal}
  {\bibinfo  {journal} {Phys. Rev. C}\ }\textbf {\bibinfo {volume} {85}},\
  \bibinfo {pages} {034002} (\bibinfo {year} {2012}{\natexlab{a}})},\ \Eprint
  {http://arxiv.org/abs/1111.3993} {arXiv:1111.3993 [nucl-th]} \BibitemShut
  {NoStop}%
\bibitem [{\citenamefont {Long}\ and\ \citenamefont
  {Yang}(2012{\natexlab{b}})}]{Long:2012ve}%
  \BibitemOpen
  \bibfield  {author} {\bibinfo {author} {\bibfnamefont {B.}~\bibnamefont
  {Long}}\ and\ \bibinfo {author} {\bibfnamefont {C.~J.}\ \bibnamefont
  {Yang}},\ }\href {\doibase 10.1103/PhysRevC.86.024001} {\bibfield  {journal}
  {\bibinfo  {journal} {Phys. Rev. C}\ }\textbf {\bibinfo {volume} {86}},\
  \bibinfo {pages} {024001} (\bibinfo {year} {2012}{\natexlab{b}})},\ \Eprint
  {http://arxiv.org/abs/1202.4053} {arXiv:1202.4053 [nucl-th]} \BibitemShut
  {NoStop}%
\bibitem [{\citenamefont {Wu}\ and\ \citenamefont {Long}(2019)}]{Wu:2018lai}%
  \BibitemOpen
  \bibfield  {author} {\bibinfo {author} {\bibfnamefont {S.}~\bibnamefont
  {Wu}}\ and\ \bibinfo {author} {\bibfnamefont {B.}~\bibnamefont {Long}},\
  }\href {\doibase 10.1103/PhysRevC.99.024003} {\bibfield  {journal} {\bibinfo
  {journal} {Phys. Rev. C}\ }\textbf {\bibinfo {volume} {99}},\ \bibinfo
  {pages} {024003} (\bibinfo {year} {2019})},\ \Eprint
  {http://arxiv.org/abs/1807.04407} {arXiv:1807.04407 [nucl-th]} \BibitemShut
  {NoStop}%
\bibitem [{\citenamefont {Pav\'on~Valderrama}\ and\ \citenamefont
  {Phillips}(2015)}]{PavonValderrama:2014zeq}%
  \BibitemOpen
  \bibfield  {author} {\bibinfo {author} {\bibfnamefont {M.}~\bibnamefont
  {Pav\'on~Valderrama}}\ and\ \bibinfo {author} {\bibfnamefont {D.~R.}\
  \bibnamefont {Phillips}},\ }\href {\doibase 10.1103/PhysRevLett.114.082502}
  {\bibfield  {journal} {\bibinfo  {journal} {Phys. Rev. Lett.}\ }\textbf
  {\bibinfo {volume} {114}},\ \bibinfo {pages} {082502} (\bibinfo {year}
  {2015})},\ \Eprint {http://arxiv.org/abs/1407.0437} {arXiv:1407.0437
  [nucl-th]} \BibitemShut {NoStop}%
\bibitem [{\citenamefont {Efros}\ \emph {et~al.}(1994)\citenamefont {Efros},
  \citenamefont {Leidemann},\ and\ \citenamefont {Orlandini}}]{Efros:1994iq}%
  \BibitemOpen
  \bibfield  {author} {\bibinfo {author} {\bibfnamefont {V.~D.}\ \bibnamefont
  {Efros}}, \bibinfo {author} {\bibfnamefont {W.}~\bibnamefont {Leidemann}}, \
  and\ \bibinfo {author} {\bibfnamefont {G.}~\bibnamefont {Orlandini}},\ }\href
  {\doibase 10.1016/0370-2693(94)91355-2} {\bibfield  {journal} {\bibinfo
  {journal} {Phys. Lett. B}\ }\textbf {\bibinfo {volume} {338}},\ \bibinfo
  {pages} {130} (\bibinfo {year} {1994})}\BibitemShut {NoStop}%
\bibitem [{\citenamefont {Efros}\ \emph {et~al.}(2007)\citenamefont {Efros},
  \citenamefont {Leidemann}, \citenamefont {Orlandini},\ and\ \citenamefont
  {Barnea}}]{Efros:2007nq}%
  \BibitemOpen
  \bibfield  {author} {\bibinfo {author} {\bibfnamefont {V.~D.}\ \bibnamefont
  {Efros}}, \bibinfo {author} {\bibfnamefont {W.}~\bibnamefont {Leidemann}},
  \bibinfo {author} {\bibfnamefont {G.}~\bibnamefont {Orlandini}}, \ and\
  \bibinfo {author} {\bibfnamefont {N.}~\bibnamefont {Barnea}},\ }\href
  {\doibase 10.1088/0954-3899/34/12/R02} {\bibfield  {journal} {\bibinfo
  {journal} {J. Phys. G: Nucl. Part. Phys.}\ }\textbf {\bibinfo {volume}
  {34}},\ \bibinfo {pages} {R459} (\bibinfo {year} {2007})},\ \bibinfo {note}
  {publisher: IOP Publishing}\BibitemShut {NoStop}%
\bibitem [{\citenamefont {Marino}\ \emph {et~al.}(2025)\citenamefont {Marino},
  \citenamefont {Bonaiti}, \citenamefont {Bacca}, \citenamefont {Hagen},\ and\
  \citenamefont {Jansen}}]{Marino:2025auh}%
  \BibitemOpen
  \bibfield  {author} {\bibinfo {author} {\bibfnamefont {F.}~\bibnamefont
  {Marino}}, \bibinfo {author} {\bibfnamefont {F.}~\bibnamefont {Bonaiti}},
  \bibinfo {author} {\bibfnamefont {S.}~\bibnamefont {Bacca}}, \bibinfo
  {author} {\bibfnamefont {G.}~\bibnamefont {Hagen}}, \ and\ \bibinfo {author}
  {\bibfnamefont {G.~R.}\ \bibnamefont {Jansen}},\ }\href@noop {} {\  (\bibinfo
  {year} {2025})},\ \Eprint {http://arxiv.org/abs/2504.11012} {arXiv:2504.11012
  [nucl-th]} \BibitemShut {NoStop}%
\bibitem [{\citenamefont {Griesshammer}\ \emph {et~al.}(2024)\citenamefont
  {Griesshammer}, \citenamefont {Liao}, \citenamefont {McGovern}, \citenamefont
  {Nogga},\ and\ \citenamefont {Phillips}}]{Griesshammer:2024twu}%
  \BibitemOpen
  \bibfield  {author} {\bibinfo {author} {\bibfnamefont {H.~W.}\ \bibnamefont
  {Griesshammer}}, \bibinfo {author} {\bibfnamefont {J.}~\bibnamefont {Liao}},
  \bibinfo {author} {\bibfnamefont {J.~A.}\ \bibnamefont {McGovern}}, \bibinfo
  {author} {\bibfnamefont {A.}~\bibnamefont {Nogga}}, \ and\ \bibinfo {author}
  {\bibfnamefont {D.~R.}\ \bibnamefont {Phillips}},\ }\href {\doibase
  10.1140/epja/s10050-024-01339-x} {\bibfield  {journal} {\bibinfo  {journal}
  {Eur. Phys. J. A}\ }\textbf {\bibinfo {volume} {60}},\ \bibinfo {pages} {132}
  (\bibinfo {year} {2024})},\ \Eprint {http://arxiv.org/abs/2401.16995}
  {arXiv:2401.16995 [nucl-th]} \BibitemShut {NoStop}%
\bibitem [{\citenamefont {de~Forest~Jr.}\ and\ \citenamefont
  {Walecka}(1966)}]{Forest:1966}%
  \BibitemOpen
  \bibfield  {author} {\bibinfo {author} {\bibfnamefont {T.}~\bibnamefont
  {de~Forest~Jr.}}\ and\ \bibinfo {author} {\bibfnamefont {J.}~\bibnamefont
  {Walecka}},\ }\href {\doibase 10.1080/00018736600101254} {\bibfield
  {journal} {\bibinfo  {journal} {Advances in Physics}\ }\textbf {\bibinfo
  {volume} {15}},\ \bibinfo {pages} {1} (\bibinfo {year} {1966})}\BibitemShut
  {NoStop}%
\bibitem [{\citenamefont {Carlson}\ and\ \citenamefont
  {Schiavilla}(1998)}]{Carlson:1997qn}%
  \BibitemOpen
  \bibfield  {author} {\bibinfo {author} {\bibfnamefont {J.}~\bibnamefont
  {Carlson}}\ and\ \bibinfo {author} {\bibfnamefont {R.}~\bibnamefont
  {Schiavilla}},\ }\href {\doibase 10.1103/RevModPhys.70.743} {\bibfield
  {journal} {\bibinfo  {journal} {Rev. Mod. Phys.}\ }\textbf {\bibinfo {volume}
  {70}},\ \bibinfo {pages} {743} (\bibinfo {year} {1998})},\ \bibinfo {note}
  {publisher: American Physical Society}\BibitemShut {NoStop}%
\bibitem [{\citenamefont {Bacca}\ and\ \citenamefont
  {Pastore}(2014)}]{Bacca:2014}%
  \BibitemOpen
  \bibfield  {author} {\bibinfo {author} {\bibfnamefont {S.}~\bibnamefont
  {Bacca}}\ and\ \bibinfo {author} {\bibfnamefont {S.}~\bibnamefont
  {Pastore}},\ }\href {\doibase 10.1088/0954-3899/41/12/123002} {\bibfield
  {journal} {\bibinfo  {journal} {Journal of Physics G: Nuclear and Particle
  Physics}\ }\textbf {\bibinfo {volume} {41}},\ \bibinfo {pages} {123002}
  (\bibinfo {year} {2014})}\BibitemShut {NoStop}%
\bibitem [{\citenamefont {Martinelli}\ \emph {et~al.}(1995)\citenamefont
  {Martinelli}, \citenamefont {Kamada}, \citenamefont {Orlandini},\ and\
  \citenamefont {Gl\"ockle}}]{Martinelli:1995vn}%
  \BibitemOpen
  \bibfield  {author} {\bibinfo {author} {\bibfnamefont {S.}~\bibnamefont
  {Martinelli}}, \bibinfo {author} {\bibfnamefont {H.}~\bibnamefont {Kamada}},
  \bibinfo {author} {\bibfnamefont {G.}~\bibnamefont {Orlandini}}, \ and\
  \bibinfo {author} {\bibfnamefont {W.}~\bibnamefont {Gl\"ockle}},\ }\href
  {\doibase 10.1103/PhysRevC.52.1778} {\bibfield  {journal} {\bibinfo
  {journal} {Phys. Rev. C}\ }\textbf {\bibinfo {volume} {52}},\ \bibinfo
  {pages} {1778} (\bibinfo {year} {1995})},\ \bibinfo {note} {publisher:
  American Physical Society}\BibitemShut {NoStop}%
\bibitem [{\citenamefont {Bacca}(2005)}]{Bacca:2005}%
  \BibitemOpen
  \bibfield  {author} {\bibinfo {author} {\bibfnamefont {S.}~\bibnamefont
  {Bacca}},\ }\emph {\bibinfo {title} {Study of electromagnetic reactions on
  light nuclei with the {Lorentz} {Integral} {Transform} method}},\ \href
  {\doibase 10.25358/OPENSCIENCE-1901} {\bibinfo {type} {Ph.{D}. {Thesis}}},\
  \bibinfo  {school} {Johannes Gutenberg-Universit{\"a}t Mainz}, \bibinfo
  {address} {Mainz} (\bibinfo {year} {2005})\BibitemShut {NoStop}%
\bibitem [{\citenamefont {Efros}\ \emph {et~al.}(1993)\citenamefont {Efros},
  \citenamefont {Leidemann},\ and\ \citenamefont {Orlandini}}]{Efros:1993xy}%
  \BibitemOpen
  \bibfield  {author} {\bibinfo {author} {\bibfnamefont {V.~D.}\ \bibnamefont
  {Efros}}, \bibinfo {author} {\bibfnamefont {W.}~\bibnamefont {Leidemann}}, \
  and\ \bibinfo {author} {\bibfnamefont {G.}~\bibnamefont {Orlandini}},\ }\href
  {\doibase 10.1007/BF01080714} {\bibfield  {journal} {\bibinfo  {journal}
  {Few-Body Syst.}\ }\textbf {\bibinfo {volume} {14}},\ \bibinfo {pages} {151}
  (\bibinfo {year} {1993})}\BibitemShut {NoStop}%
\bibitem [{\citenamefont {Epelbaum}\ \emph {et~al.}(2000)\citenamefont
  {Epelbaum}, \citenamefont {Gloeckle},\ and\ \citenamefont
  {Meissner}}]{Epelbaum:1999dj}%
  \BibitemOpen
  \bibfield  {author} {\bibinfo {author} {\bibfnamefont {E.}~\bibnamefont
  {Epelbaum}}, \bibinfo {author} {\bibfnamefont {W.}~\bibnamefont {Gloeckle}},
  \ and\ \bibinfo {author} {\bibfnamefont {U.-G.}\ \bibnamefont {Meissner}},\
  }\href {\doibase 10.1016/S0375-9474(99)00821-0} {\bibfield  {journal}
  {\bibinfo  {journal} {Nucl. Phys. A}\ }\textbf {\bibinfo {volume} {671}},\
  \bibinfo {pages} {295} (\bibinfo {year} {2000})},\ \Eprint
  {http://arxiv.org/abs/nucl-th/9910064} {arXiv:nucl-th/9910064} \BibitemShut
  {NoStop}%
\bibitem [{\citenamefont {Kong}\ and\ \citenamefont
  {Ravndal}(2000)}]{Kong:1999sf}%
  \BibitemOpen
  \bibfield  {author} {\bibinfo {author} {\bibfnamefont {X.}~\bibnamefont
  {Kong}}\ and\ \bibinfo {author} {\bibfnamefont {F.}~\bibnamefont {Ravndal}},\
  }\href {\doibase 10.1016/S0375-9474(99)00406-6} {\bibfield  {journal}
  {\bibinfo  {journal} {Nucl. Phys. A}\ }\textbf {\bibinfo {volume} {665}},\
  \bibinfo {pages} {137} (\bibinfo {year} {2000})},\ \Eprint
  {http://arxiv.org/abs/hep-ph/9903523} {arXiv:hep-ph/9903523} \BibitemShut
  {NoStop}%
\bibitem [{\citenamefont {Kaplan}(2020)}]{Kaplan:2019znu}%
  \BibitemOpen
  \bibfield  {author} {\bibinfo {author} {\bibfnamefont {D.~B.}\ \bibnamefont
  {Kaplan}},\ }\href {\doibase 10.1103/PhysRevC.102.034004} {\bibfield
  {journal} {\bibinfo  {journal} {Phys. Rev. C}\ }\textbf {\bibinfo {volume}
  {102}},\ \bibinfo {pages} {034004} (\bibinfo {year} {2020})},\ \Eprint
  {http://arxiv.org/abs/1905.07485} {arXiv:1905.07485 [nucl-th]} \BibitemShut
  {NoStop}%
\bibitem [{\citenamefont {Frank}\ \emph {et~al.}(1971)\citenamefont {Frank},
  \citenamefont {Land},\ and\ \citenamefont {Spector}}]{Frank:1971xx}%
  \BibitemOpen
  \bibfield  {author} {\bibinfo {author} {\bibfnamefont {W.}~\bibnamefont
  {Frank}}, \bibinfo {author} {\bibfnamefont {D.~J.}\ \bibnamefont {Land}}, \
  and\ \bibinfo {author} {\bibfnamefont {R.~M.}\ \bibnamefont {Spector}},\
  }\href {\doibase 10.1103/RevModPhys.43.36} {\bibfield  {journal} {\bibinfo
  {journal} {Rev. Mod. Phys.}\ }\textbf {\bibinfo {volume} {43}},\ \bibinfo
  {pages} {36} (\bibinfo {year} {1971})}\BibitemShut {NoStop}%
\bibitem [{NNo()}]{NNonline}%
  \BibitemOpen
  \href@noop {} {\enquote {\bibinfo {title} {The nn-online},}\ }\bibinfo
  {howpublished} {\url{http://nn-online.org}}\BibitemShut {NoStop}%
\bibitem [{\citenamefont {Stoks}\ \emph {et~al.}(1993)\citenamefont {Stoks},
  \citenamefont {Klomp}, \citenamefont {Rentmeester},\ and\ \citenamefont
  {de~Swart}}]{Stoks:1993tb}%
  \BibitemOpen
  \bibfield  {author} {\bibinfo {author} {\bibfnamefont {V.~G.~J.}\
  \bibnamefont {Stoks}}, \bibinfo {author} {\bibfnamefont {R.~A.~M.}\
  \bibnamefont {Klomp}}, \bibinfo {author} {\bibfnamefont {M.~C.~M.}\
  \bibnamefont {Rentmeester}}, \ and\ \bibinfo {author} {\bibfnamefont {J.~J.}\
  \bibnamefont {de~Swart}},\ }\href {\doibase 10.1103/PhysRevC.48.792}
  {\bibfield  {journal} {\bibinfo  {journal} {Phys. Rev. C}\ }\textbf {\bibinfo
  {volume} {48}},\ \bibinfo {pages} {792} (\bibinfo {year} {1993})}\BibitemShut
  {NoStop}%
\bibitem [{\citenamefont {Pav{\'o}n~Valderrama}\ \emph
  {et~al.}(2017)\citenamefont {Pav{\'o}n~Valderrama}, \citenamefont
  {S{\'a}nchez~S{\'a}nchez}, \citenamefont {Yang}, \citenamefont {Long},
  \citenamefont {Carbonell},\ and\ \citenamefont {van
  Kolck}}]{PavonValderrama:2016lqn}%
  \BibitemOpen
  \bibfield  {author} {\bibinfo {author} {\bibfnamefont {M.}~\bibnamefont
  {Pav{\'o}n~Valderrama}}, \bibinfo {author} {\bibfnamefont {M.}~\bibnamefont
  {S{\'a}nchez~S{\'a}nchez}}, \bibinfo {author} {\bibfnamefont {C.~J.}\
  \bibnamefont {Yang}}, \bibinfo {author} {\bibfnamefont {B.}~\bibnamefont
  {Long}}, \bibinfo {author} {\bibfnamefont {J.}~\bibnamefont {Carbonell}}, \
  and\ \bibinfo {author} {\bibfnamefont {U.}~\bibnamefont {van Kolck}},\ }\href
  {\doibase 10.1103/PhysRevC.95.054001} {\bibfield  {journal} {\bibinfo
  {journal} {Phys. Rev. C}\ }\textbf {\bibinfo {volume} {95}},\ \bibinfo
  {pages} {054001} (\bibinfo {year} {2017})},\ \Eprint
  {http://arxiv.org/abs/1611.10175} {arXiv:1611.10175 [nucl-th]} \BibitemShut
  {NoStop}%
\bibitem [{\citenamefont {Phillips}(2003)}]{Phillips:2003jz}%
  \BibitemOpen
  \bibfield  {author} {\bibinfo {author} {\bibfnamefont {D.~R.}\ \bibnamefont
  {Phillips}},\ }\href {\doibase 10.1016/S0370-2693(03)00867-0} {\bibfield
  {journal} {\bibinfo  {journal} {Phys. Lett. B}\ }\textbf {\bibinfo {volume}
  {567}},\ \bibinfo {pages} {12} (\bibinfo {year} {2003})}\BibitemShut
  {NoStop}%
\bibitem [{\citenamefont {Krebs}\ \emph {et~al.}(2019)\citenamefont {Krebs},
  \citenamefont {Epelbaum},\ and\ \citenamefont {Mei{\ss}ner}}]{Krebs:2019aka}%
  \BibitemOpen
  \bibfield  {author} {\bibinfo {author} {\bibfnamefont {H.}~\bibnamefont
  {Krebs}}, \bibinfo {author} {\bibfnamefont {E.}~\bibnamefont {Epelbaum}}, \
  and\ \bibinfo {author} {\bibfnamefont {U.-G.}\ \bibnamefont {Mei{\ss}ner}},\
  }\href {\doibase 10.1007/s00601-019-1500-5} {\bibfield  {journal} {\bibinfo
  {journal} {Few-Body Syst.}\ }\textbf {\bibinfo {volume} {60}},\ \bibinfo
  {pages} {31} (\bibinfo {year} {2019})}\BibitemShut {NoStop}%
\bibitem [{\citenamefont {Schiavilla}\ and\ \citenamefont
  {Pandharipande}(2002)}]{Schiavilla:2002fq}%
  \BibitemOpen
  \bibfield  {author} {\bibinfo {author} {\bibfnamefont {R.}~\bibnamefont
  {Schiavilla}}\ and\ \bibinfo {author} {\bibfnamefont {V.~R.}\ \bibnamefont
  {Pandharipande}},\ }\href {\doibase 10.1103/PhysRevC.65.064009} {\bibfield
  {journal} {\bibinfo  {journal} {Phys. Rev. C}\ }\textbf {\bibinfo {volume}
  {65}},\ \bibinfo {pages} {064009} (\bibinfo {year} {2002})},\ \Eprint
  {http://arxiv.org/abs/nucl-th/0201043} {arXiv:nucl-th/0201043} \BibitemShut
  {NoStop}%
\bibitem [{\citenamefont {More}\ \emph {et~al.}(2015)\citenamefont {More},
  \citenamefont {K{\"o}nig}, \citenamefont {Furnstahl},\ and\ \citenamefont
  {Hebeler}}]{More:2015tpa}%
  \BibitemOpen
  \bibfield  {author} {\bibinfo {author} {\bibfnamefont {S.~N.}\ \bibnamefont
  {More}}, \bibinfo {author} {\bibfnamefont {S.}~\bibnamefont {K{\"o}nig}},
  \bibinfo {author} {\bibfnamefont {R.~J.}\ \bibnamefont {Furnstahl}}, \ and\
  \bibinfo {author} {\bibfnamefont {K.}~\bibnamefont {Hebeler}},\ }\href
  {\doibase 10.1103/PhysRevC.92.064002} {\bibfield  {journal} {\bibinfo
  {journal} {Phys. Rev. C}\ }\textbf {\bibinfo {volume} {92}},\ \bibinfo
  {pages} {064002} (\bibinfo {year} {2015})},\ \Eprint
  {http://arxiv.org/abs/1510.04955} {arXiv:1510.04955 [nucl-th]} \BibitemShut
  {NoStop}%
\bibitem [{\citenamefont {More}\ \emph {et~al.}(2017)\citenamefont {More},
  \citenamefont {Bogner},\ and\ \citenamefont {Furnstahl}}]{More:2017syr}%
  \BibitemOpen
  \bibfield  {author} {\bibinfo {author} {\bibfnamefont {S.~N.}\ \bibnamefont
  {More}}, \bibinfo {author} {\bibfnamefont {S.~K.}\ \bibnamefont {Bogner}}, \
  and\ \bibinfo {author} {\bibfnamefont {R.~J.}\ \bibnamefont {Furnstahl}},\
  }\href {\doibase 10.1103/PhysRevC.96.054004} {\bibfield  {journal} {\bibinfo
  {journal} {Phys. Rev. C}\ }\textbf {\bibinfo {volume} {96}},\ \bibinfo
  {pages} {054004} (\bibinfo {year} {2017})},\ \Eprint
  {http://arxiv.org/abs/1708.03315} {arXiv:1708.03315 [nucl-th]} \BibitemShut
  {NoStop}%
\bibitem [{\citenamefont {Simon}\ \emph {et~al.}(1979)\citenamefont {Simon},
  \citenamefont {Borkowski}, \citenamefont {Schmitt}, \citenamefont {Walther},
  \citenamefont {Arenh{\"o}vel},\ and\ \citenamefont {Fabian}}]{Simon:1979bu}%
  \BibitemOpen
  \bibfield  {author} {\bibinfo {author} {\bibfnamefont {G.~G.}\ \bibnamefont
  {Simon}}, \bibinfo {author} {\bibfnamefont {F.}~\bibnamefont {Borkowski}},
  \bibinfo {author} {\bibfnamefont {{\relax Ch}.}~\bibnamefont {Schmitt}},
  \bibinfo {author} {\bibfnamefont {V.~H.}\ \bibnamefont {Walther}}, \bibinfo
  {author} {\bibfnamefont {H.}~\bibnamefont {Arenh{\"o}vel}}, \ and\ \bibinfo
  {author} {\bibfnamefont {W.}~\bibnamefont {Fabian}},\ }\href {\doibase
  10.1016/0375-9474(79)90584-0} {\bibfield  {journal} {\bibinfo  {journal}
  {Nucl. Phys. A}\ }\textbf {\bibinfo {volume} {324}},\ \bibinfo {pages} {277}
  (\bibinfo {year} {1979})}\BibitemShut {NoStop}%
\bibitem [{\citenamefont {Fabian}\ and\ \citenamefont
  {Arenh{\"o}vel}(1979)}]{Fabian:1979kx}%
  \BibitemOpen
  \bibfield  {author} {\bibinfo {author} {\bibfnamefont {W.}~\bibnamefont
  {Fabian}}\ and\ \bibinfo {author} {\bibfnamefont {H.}~\bibnamefont
  {Arenh{\"o}vel}},\ }\href {\doibase 10.1016/0375-9474(79)90599-2} {\bibfield
  {journal} {\bibinfo  {journal} {Nucl. Phys. A}\ }\textbf {\bibinfo {volume}
  {314}},\ \bibinfo {pages} {253} (\bibinfo {year} {1979})}\BibitemShut
  {NoStop}%
\bibitem [{\citenamefont {Arenh{\"o}vel}\ \emph {et~al.}(2005)\citenamefont
  {Arenh{\"o}vel}, \citenamefont {Leidemann},\ and\ \citenamefont
  {Tomusiak}}]{Arenhovel:2004bc}%
  \BibitemOpen
  \bibfield  {author} {\bibinfo {author} {\bibfnamefont {H.}~\bibnamefont
  {Arenh{\"o}vel}}, \bibinfo {author} {\bibfnamefont {W.}~\bibnamefont
  {Leidemann}}, \ and\ \bibinfo {author} {\bibfnamefont {E.~L.}\ \bibnamefont
  {Tomusiak}},\ }\href {\doibase 10.1140/epja/i2004-10061-5} {\bibfield
  {journal} {\bibinfo  {journal} {Eur. Phys. J. A}\ }\textbf {\bibinfo {volume}
  {23}},\ \bibinfo {pages} {147} (\bibinfo {year} {2005})}\BibitemShut
  {NoStop}%
\bibitem [{\citenamefont {Christlmeier}\ and\ \citenamefont
  {Griesshammer}(2008)}]{Christlmeier:2008ye}%
  \BibitemOpen
  \bibfield  {author} {\bibinfo {author} {\bibfnamefont {S.}~\bibnamefont
  {Christlmeier}}\ and\ \bibinfo {author} {\bibfnamefont {H.~W.}\ \bibnamefont
  {Griesshammer}},\ }\href {\doibase 10.1103/PhysRevC.77.064001} {\bibfield
  {journal} {\bibinfo  {journal} {Phys. Rev. C}\ }\textbf {\bibinfo {volume}
  {77}},\ \bibinfo {pages} {064001} (\bibinfo {year} {2008})},\ \Eprint
  {http://arxiv.org/abs/0803.1307} {arXiv:0803.1307 [nucl-th]} \BibitemShut
  {NoStop}%
\bibitem [{\citenamefont {Teng}\ and\ \citenamefont
  {Griesshammer}(2025)}]{Teng:2024exc}%
  \BibitemOpen
  \bibfield  {author} {\bibinfo {author} {\bibfnamefont {Y.~P.}\ \bibnamefont
  {Teng}}\ and\ \bibinfo {author} {\bibfnamefont {H.~W.}\ \bibnamefont
  {Griesshammer}},\ }\href {\doibase 10.1140/epja/s10050-025-01675-6}
  {\bibfield  {journal} {\bibinfo  {journal} {Eur. Phys. J. A}\ }\textbf
  {\bibinfo {volume} {61}},\ \bibinfo {pages} {211} (\bibinfo {year} {2025})},\
  \Eprint {http://arxiv.org/abs/2410.09653} {arXiv:2410.09653 [nucl-th]}
  \BibitemShut {NoStop}%
\bibitem [{\citenamefont {Lyu}\ \emph {et~al.}(2025)\citenamefont {Lyu},
  \citenamefont {Zuo}, \citenamefont {Peng}, \citenamefont {K{\"o}nig},\ and\
  \citenamefont {Long}}]{Lyu:2025yhz}%
  \BibitemOpen
  \bibfield  {author} {\bibinfo {author} {\bibfnamefont {S.}~\bibnamefont
  {Lyu}}, \bibinfo {author} {\bibfnamefont {L.}~\bibnamefont {Zuo}}, \bibinfo
  {author} {\bibfnamefont {R.}~\bibnamefont {Peng}}, \bibinfo {author}
  {\bibfnamefont {S.}~\bibnamefont {K{\"o}nig}}, \ and\ \bibinfo {author}
  {\bibfnamefont {B.}~\bibnamefont {Long}},\ }\href@noop {} {\  (\bibinfo
  {year} {2025})},\ \Eprint {http://arxiv.org/abs/2511.12522} {arXiv:2511.12522
  [nucl-th]} \BibitemShut {NoStop}%
\bibitem [{\citenamefont {K{\"o}nig}(2020)}]{Konig:2019xxk}%
  \BibitemOpen
  \bibfield  {author} {\bibinfo {author} {\bibfnamefont {S.}~\bibnamefont
  {K{\"o}nig}},\ }\href {\doibase 10.1140/epja/s10050-020-00098-9} {\bibfield
  {journal} {\bibinfo  {journal} {Eur. Phys. J. A}\ }\textbf {\bibinfo {volume}
  {56}},\ \bibinfo {pages} {113} (\bibinfo {year} {2020})}\BibitemShut
  {NoStop}%
\bibitem [{\citenamefont {Peng}\ \emph {et~al.}(2022)\citenamefont {Peng},
  \citenamefont {Lyu}, \citenamefont {K{\"o}nig},\ and\ \citenamefont
  {Long}}]{Peng:2021pvo}%
  \BibitemOpen
  \bibfield  {author} {\bibinfo {author} {\bibfnamefont {R.}~\bibnamefont
  {Peng}}, \bibinfo {author} {\bibfnamefont {S.}~\bibnamefont {Lyu}}, \bibinfo
  {author} {\bibfnamefont {S.}~\bibnamefont {K{\"o}nig}}, \ and\ \bibinfo
  {author} {\bibfnamefont {B.}~\bibnamefont {Long}},\ }\href {\doibase
  10.1103/PhysRevC.105.054002} {\bibfield  {journal} {\bibinfo  {journal}
  {Phys. Rev. C}\ }\textbf {\bibinfo {volume} {105}},\ \bibinfo {pages}
  {054002} (\bibinfo {year} {2022})},\ \Eprint
  {http://arxiv.org/abs/2112.00947} {arXiv:2112.00947 [nucl-th]} \BibitemShut
  {NoStop}%
\bibitem [{\citenamefont {Bacca}\ \emph {et~al.}(2013)\citenamefont {Bacca},
  \citenamefont {Barnea}, \citenamefont {Leidemann},\ and\ \citenamefont
  {Orlandini}}]{Bacca:2012xv}%
  \BibitemOpen
  \bibfield  {author} {\bibinfo {author} {\bibfnamefont {S.}~\bibnamefont
  {Bacca}}, \bibinfo {author} {\bibfnamefont {N.}~\bibnamefont {Barnea}},
  \bibinfo {author} {\bibfnamefont {W.}~\bibnamefont {Leidemann}}, \ and\
  \bibinfo {author} {\bibfnamefont {G.}~\bibnamefont {Orlandini}},\ }\href
  {\doibase 10.1103/PhysRevLett.110.042503} {\bibfield  {journal} {\bibinfo
  {journal} {Phys. Rev. Lett.}\ }\textbf {\bibinfo {volume} {110}},\ \bibinfo
  {pages} {042503} (\bibinfo {year} {2013})},\ \bibinfo {note} {publisher:
  American Physical Society}\BibitemShut {NoStop}%
\bibitem [{\citenamefont {Kegel}\ \emph {et~al.}(2023)\citenamefont {Kegel}
  \emph {et~al.}}]{Kegel:2021jrh}%
  \BibitemOpen
  \bibfield  {author} {\bibinfo {author} {\bibfnamefont {S.}~\bibnamefont
  {Kegel}} \emph {et~al.},\ }\href {\doibase 10.1103/PhysRevLett.130.152502}
  {\bibfield  {journal} {\bibinfo  {journal} {Phys. Rev. Lett.}\ }\textbf
  {\bibinfo {volume} {130}},\ \bibinfo {pages} {152502} (\bibinfo {year}
  {2023})},\ \Eprint {http://arxiv.org/abs/2112.10582} {arXiv:2112.10582
  [nucl-ex]} \BibitemShut {NoStop}%
\bibitem [{\citenamefont {Michel}\ \emph {et~al.}(2023)\citenamefont {Michel},
  \citenamefont {Nazarewicz},\ and\ \citenamefont
  {P{\l}oszajczak}}]{Michel:2023ley}%
  \BibitemOpen
  \bibfield  {author} {\bibinfo {author} {\bibfnamefont {N.}~\bibnamefont
  {Michel}}, \bibinfo {author} {\bibfnamefont {W.}~\bibnamefont {Nazarewicz}},
  \ and\ \bibinfo {author} {\bibfnamefont {M.}~\bibnamefont {P{\l}oszajczak}},\
  }\href {\doibase 10.1103/PhysRevLett.131.242502} {\bibfield  {journal}
  {\bibinfo  {journal} {Phys. Rev. Lett.}\ }\textbf {\bibinfo {volume} {131}},\
  \bibinfo {pages} {242502} (\bibinfo {year} {2023})},\ \bibinfo {note}
  {[Erratum: Phys.Rev.Lett. 133, 239901 (2024)]},\ \Eprint
  {http://arxiv.org/abs/2306.05192} {arXiv:2306.05192 [nucl-th]} \BibitemShut
  {NoStop}%
\bibitem [{\citenamefont {Mei{\ss}ner}\ \emph {et~al.}(2024)\citenamefont
  {Mei{\ss}ner}, \citenamefont {Shen}, \citenamefont {Elhatisari},\ and\
  \citenamefont {Lee}}]{Meissner:2023cvo}%
  \BibitemOpen
  \bibfield  {author} {\bibinfo {author} {\bibfnamefont {U.-G.}\ \bibnamefont
  {Mei{\ss}ner}}, \bibinfo {author} {\bibfnamefont {S.}~\bibnamefont {Shen}},
  \bibinfo {author} {\bibfnamefont {S.}~\bibnamefont {Elhatisari}}, \ and\
  \bibinfo {author} {\bibfnamefont {D.}~\bibnamefont {Lee}},\ }\href {\doibase
  10.1103/PhysRevLett.132.062501} {\bibfield  {journal} {\bibinfo  {journal}
  {Phys. Rev. Lett.}\ }\textbf {\bibinfo {volume} {132}},\ \bibinfo {pages}
  {062501} (\bibinfo {year} {2024})},\ \Eprint
  {http://arxiv.org/abs/2309.01558} {arXiv:2309.01558 [nucl-th]} \BibitemShut
  {NoStop}%
\bibitem [{\citenamefont {Yin}\ \emph {et~al.}(2025)\citenamefont {Yin},
  \citenamefont {Shirokov}, \citenamefont {Li}, \citenamefont {Zhou},
  \citenamefont {Zhao}, \citenamefont {Bacca},\ and\ \citenamefont
  {Vary}}]{Yin:2024xsg}%
  \BibitemOpen
  \bibfield  {author} {\bibinfo {author} {\bibfnamefont {P.}~\bibnamefont
  {Yin}}, \bibinfo {author} {\bibfnamefont {A.~M.}\ \bibnamefont {Shirokov}},
  \bibinfo {author} {\bibfnamefont {H.}~\bibnamefont {Li}}, \bibinfo {author}
  {\bibfnamefont {B.}~\bibnamefont {Zhou}}, \bibinfo {author} {\bibfnamefont
  {X.}~\bibnamefont {Zhao}}, \bibinfo {author} {\bibfnamefont {S.}~\bibnamefont
  {Bacca}}, \ and\ \bibinfo {author} {\bibfnamefont {J.~P.}\ \bibnamefont
  {Vary}},\ }\href {\doibase 10.1103/1pxs-xjtl} {\bibfield  {journal} {\bibinfo
   {journal} {Phys. Rev. C}\ }\textbf {\bibinfo {volume} {112}},\ \bibinfo
  {pages} {L031303} (\bibinfo {year} {2025})},\ \Eprint
  {http://arxiv.org/abs/2412.18037} {arXiv:2412.18037 [nucl-th]} \BibitemShut
  {NoStop}%
\bibitem [{\citenamefont {Gl{\"o}ckle}(1983)}]{Gloeckle:1983}%
  \BibitemOpen
  \bibfield  {author} {\bibinfo {author} {\bibfnamefont {W.}~\bibnamefont
  {Gl{\"o}ckle}},\ }\href@noop {} {\emph {\bibinfo {title} {The Quantum
  Mechanical Few-Body Problem}}}\ (\bibinfo  {publisher} {Springer},\ \bibinfo
  {address} {Berlin; New York},\ \bibinfo {year} {1983})\BibitemShut {NoStop}%
\bibitem [{\citenamefont {Varshalovich}\ \emph {et~al.}(1988)\citenamefont
  {Varshalovich}, \citenamefont {Moskalev},\ and\ \citenamefont
  {Khersonskii}}]{Varshalovich:1988ifq}%
  \BibitemOpen
  \bibfield  {author} {\bibinfo {author} {\bibfnamefont {D.~A.}\ \bibnamefont
  {Varshalovich}}, \bibinfo {author} {\bibfnamefont {A.~N.}\ \bibnamefont
  {Moskalev}}, \ and\ \bibinfo {author} {\bibfnamefont {V.~K.}\ \bibnamefont
  {Khersonskii}},\ }\href {\doibase 10.1142/0270} {\emph {\bibinfo {title}
  {{Quantum Theory of Angular Momentum}: {Irreducible Tensors, Spherical
  Harmonics, Vector Coupling Coefficients, 3nj Symbols}}}}\ (\bibinfo
  {publisher} {World Scientific Publishing Company},\ \bibinfo {year}
  {1988})\BibitemShut {NoStop}%
\bibitem [{\citenamefont {Hjorth-Jensen}(2023)}]{HjorthJensen:2023}%
  \BibitemOpen
  \bibfield  {author} {\bibinfo {author} {\bibfnamefont {M.}~\bibnamefont
  {Hjorth-Jensen}},\ }\href@noop {} {\enquote {\bibinfo {title} {Applied data
  analysis and machine learning},}\ }\bibinfo {howpublished}
  {\url{https://compphysics.github.io/MachineLearning/doc/LectureNotes/}}
  (\bibinfo {year} {2023})\BibitemShut {NoStop}%
\end{thebibliography}

\end{document}